\documentclass[useAMS,usenatbib]{mn2e}
\usepackage{epsfig}
\usepackage{verbatim} 
\usepackage{fixltx2e}

\def\msun{{\rm {M}}_{\odot}}
\newcommand{\etal}{{et al.}~}

\newcommand{\eg}{{e.g.~}}
\newcommand{\ie}{{i.e.~}}

\def \ltsima{$\; \buildrel < \over \sim \;$}
\def \simlt{\lower.5ex\hbox{\ltsima}}            
\def \gtsima{$\; \buildrel > \over \sim \;$}
\def \gtsima{\mbox{$\; \buildrel > \over \sim \;$}}
\def \simgt{\lower.5ex\hbox{\gtsima}}            

\newcounter{cureqno}

\title[halo universality]{Non-universality of halo profiles and implications for dark matter experiments
}

\author[Reed \etal] {
Darren S. Reed$^{1}$\thanks{email: reed@physik.uzh.ch},
Savvas M. Koushiappas$^{2}$, 
Liang Gao$^{3,4}$\\
$^1$Insitute for Theoretical Physics, Univ. of Z\"{u}rich, Winterthurerstrasse 190, CH-8057 Z\"{u}rich, Switzerland\\
$^2$Department of Physics, Brown University, 182 Hope St., Providence, RI 02912, U.S.A.\\
$^3$National Astronomical Observationaries, Chinese Academy of Science,
Beijing, 100012, China \\
$^4$Institute for Computational Cosmology, Dept. of Physics, University of Durham, South Road, Durham DH1 3LE, UK\\
}
\pagerange{\pageref{firstpage}--\pageref{lastpage}}
\pubyear{xxxx}

\begin{document}

\maketitle

\label{firstpage}

\begin{abstract}

We explore the cosmological halo-to-halo scatter of the distribution
of mass within dark matter halos utilizing a well-resolved statistical
sample of clusters from the cosmological Millennium simulation.  We
find that at any radius, the spherically-averaged dark matter density
of a halo (corresponding to the ``smooth-component'') and its
logarithmic slope are well-described by a Gaussian probability
distribution.  At small radii (within the scale radius), the density
distribution is fully determined by the measured Gaussian distribution
in halo concentrations.  The variance in the radial distribution of
mass in dark matter halos is important for the interpretation of
direct and indirect dark matter detection efforts.  The scatter in
mass profiles imparts approximately a 25 percent cosmological
uncertainty in the dark matter density at the Solar neighborhood and a
factor of $\sim$3 uncertainty in the expected Galactic dark matter
annihilation flux.  The aggregate effect of halo-to-halo profile
scatter leads to a small (few percent) enhancement in dark matter
annihilation background if the Gaussian concentration distribution
holds for all halo masses versus a 10 percent enhancement under the
assumption of a log-normal concentration distribution.  The Gaussian
nature of the cluster profile scatter implies that the technique of
``stacking'' halos to improve signal to noise should not suffer from
bias.

\end{abstract}

\section{introduction}

Overwhelming evidence indicates that most of the mass in the Universe
is composed of dark matter.  In the concordance $\Lambda$ Cold Dark
Matter ($\Lambda$CDM) cosmological model, structure formation is
dominated by the gravitational evolution of dark matter.  At the
present epoch, most of the mass has been assembled into self-bound
halos, which are hosts to the galaxies, clusters, and groups that are
observed.  Although the exact nature of dark matter is unknown,
theoretically motivated extensions to the standard model of particle
physics suggest cold dark matter candidates which were in thermal
equilibrium in the early Universe and interact only weakly with
baryonic matter.


Weakly Interacting Massive Particles (WIMPs) are ideal dark matter
candidates as they arise naturally in many extensions of the standard
model of particle physics. The strength of their interactions can
mimic the physical behaviour of the dark matter (weak as well as
gravitational) inferred from a broad range of observations.  WIMP dark
matter candidates have a small but non-zero cross-section for
self-annihilation (\citealt{jungman}). Indirect detection experiments
look for the by-products of this annihilation, typically in the form
of high-energy photons, neutrinos and positrons, as well as low-energy
antiprotons (see \citealt{jungman}; \citealt{bertone}).

In recent years, halo structure has been widely explored using
cosmological numerical simulations.  Pioneering work by Navarro, Frenk
\& White (1996, 1997; NFW hereafter) used numerical simulations to
show that the spherically-averaged radial density profile of dark
matter halos is approximately ``universal''.  However, NFW and later
\cite{jing} and \cite{bullock} noted significant variations in the
profile between different halos, in that some are better fit by
steeper (higher concentration) profile forms than others.  These
variations were shown to be correlated with halo formation time (\eg
\citealt{wechsler}).  Upon close inspection, the profile of any
individual halo cannot be described by any particular smooth
functional form (\eg \citealt{jing}; \citealt{reedpro};
\citealt{gaoasym}; \citealt{knollmann}; \citealt{lukic09}).  In this
sense, the halo mass profile is not truly ``universal''.  Thus, due to
this fundamentally non-smooth nature of the radial mass distribution
of halos, functional forms do not provide a complete and accurate
description of the halo density.

In this paper, we quantify the mean and scatter of the halo radial
mass distribution without the {\it a priori} assumption of a smooth
functional form.  We use a large sample of halos extracted from the
``Millennium'' cosmological numerical simulation, whose combination of
large volume and fine mass resolution are ideal for studying
statistical variations in halo mass profiles.  Our approach begins
with an empirical non-parametric measure of the distribution function
of halo densities, allowing us to include the effects of halo-to-halo
and intra-halo non-universality in addition to effects relating to
scatter in halo concentrations.  We focus on small radii where the
``smooth'' dark matter component dominates the mass distribution;
here, self-bound satellite ``subhalos'' are deficient because of
efficient tidal stripping before reaching small radii
(\citealt{springelaq}).  This allows us to quantify the cosmological
scatter in the smooth component of the dark matter density.  We apply
our results to several cosmological applications, including direct and
indirect experimental efforts of dark matter detection.

Annihilation of dark matter scales as the square of the number density
of particles, and thus any detected annihilation signal will be
sensitive to the precise distribution of dark matter. Past work
focused on using extremely high resolution simulations of individual
dark matter halos of Galactic mass to estimate annihilation
luminosities for particular dark matter candidates
(\citealt{kuhlendm}; \citealt{springeldm}). These impressive numerical
simulations were able to quantify the level at which substructure
contributes to the annihilation signal, as well as get a glimpse of
the phase-space structure of the Milky Way halo at the Solar
neighbourhood.  However, halo-to-halo variations in the radial mass
profile implies a ``cosmological'' uncertainty in the predicted
annihilation rate in halos, which is of course in addition to the
uncertainties related to the mass, cross-section, and other properties
of the dark matter particle.

Dark matter annihilation in halos produces a cosmological background
whose strength depends upon the numbers of halos throughout the
history of the Universe and their density profiles (\eg
\citealt{ullio}; \citealt{zavala}).  When integrated over all halos, a
scatter in halo density profiles implies a boost of the annihilation
background with respect to the case where all halos follow a universal
profile without scatter (\eg \citealt{ullio}).  The strength of the
annihilation background will thus be sensitive to both the average
{\it and} the scatter in halo dark matter profiles.  Inferring the
dark matter particle mass and cross section from a background
annihilation signal will thus require separating out the integrated
effects of cosmological scatter.

In addition, even within the Milky Way, interpretation of both direct
and indirect dark matter detection efforts requires an understanding
of the intrinsic variance in halo profiles.  The cosmological
uncertainty in the local dark matter density at the solar radius is
essential in the interpretation of a detection (or lack of) in direct
detection experiments.  In particular, knowledge of the intra-halo
scatter in density with radius is needed to evaluate results of direct
(local) dark matter detection efforts in the context of indirect
(non-local) detection experiments focussed on the Galactic center or
elsewhere.  Thus, both direct and indirect dark matter detection
efforts face the challenge of disentangling the influences of
cosmological halos from the properties of the dark matter particle.

In this paper, we use a large number of cosmological halos to quantify
the cosmological variance in halo densities, and to assess the level
at which the distribution of halo densities affects the interpretation
of dark matter direct and indirect detection experiments.  In
\S~\ref{sec:haloprofile}, we review the general characteristics of
dark matter halo mass distribution, and we give an overview of the
cosmological simulations that we use in \S~\ref{sec:thesimulation}.
We present our analysis of the mass distribution and its scatter
within halos in the simulation in \S~\ref{sec:resultsprofile}, showing
that the distribution of halo concentrations yields an accurate
description of the measured scatter of the smooth component of dark
matter within halos. We apply our findings to experimental searches of
dark matter in \S~\ref{sec:annil}; we discuss limitations of our work
(\S~\ref{sec:limitations}), followed by a brief conclusion
(\S~\ref{sec:conclusions}).

\section{The dark matter halo profile}
\label{sec:haloprofile}

As a baseline reference for examining the cosmological scatter of halo
density profiles, it is convenient to parameterize the density profile
by a spherically-averaged functional form.  Recent works favor the
\cite{einasto} profile form as a description of cosmological halos
(\citealt{juliopro}; \citealt{gaoconcs}; \citealt{hayashiwhite}).  In
this case, the logarithmic slope of the density is a simple power-law:
\begin{equation}
{d\ln\rho \over d\ln r}=-2 \left({r \over r_{-2}}\right)^{\alpha}.
\label{eqnein}
\end{equation}
The ``scale radius'', $r=r_{-2}$, has a density slope of $-2$, and
defines the halo concentration as:
\begin{equation}
{c_{vir}=\frac{r_{vir}}{r_{-2}}, ~{\rm or} ~c_{200}=\frac{r_{200}}{r_{-2}}}, 
\end{equation}
where in the first definition, $r_{vir}$ is the virial radius of the
halo, defined as a sphere of 95.4 times critical density
(\citealt{eke}), while in the second definition, $r_{200}$ is the
radius where the enclosed density is 200 times the critical density of
the Universe.  The parameter $\alpha$ in Eq.~\ref{eqnein} varies
weakly with mass and redshift.  On average $\alpha=0.19$, for $z=0$
clusters over the mass range that we explore (as shown in
\citealt{gaoconcs}).  Density is given by:
\begin{equation} 
\rho(r) = \rho_{-2} \exp \left\{ - \frac{ 2}{\alpha} \left[ \left( \frac{
          r}{r_{-2}} \right)^\alpha - 1 \right] \right\}.
\label{eqneinrho}
\end{equation}
The normalization of the profile is obtained from requiring that the
mass of the halo is $M = \int_0^{R_{200}} 4 \pi \rho(r) r^2 dr$, thus
\begin{equation} 
\rho_{-2} = \frac{ 2^{3/ \alpha} c_{200}^3 M}{4 \pi\, e^{2/\alpha}
  \alpha^{(3-\alpha)/\alpha} R_{200}^3 \left[ \Gamma \left(
    \frac{3}{\alpha} \right) - \Gamma \left( \frac{3}{\alpha},
    \frac{2c_{200}^\alpha}{\alpha} \right) \right] }.
\label{eq:normalization}
\end{equation}
Here, $\Gamma(x)$ and $\Gamma(x,y)$ are the Gamma, and Incomplete
Gamma functions respectively.

\begin{figure*}
\centering
\begin{tabular}{cc}
  \includegraphics[width=.5\textwidth]{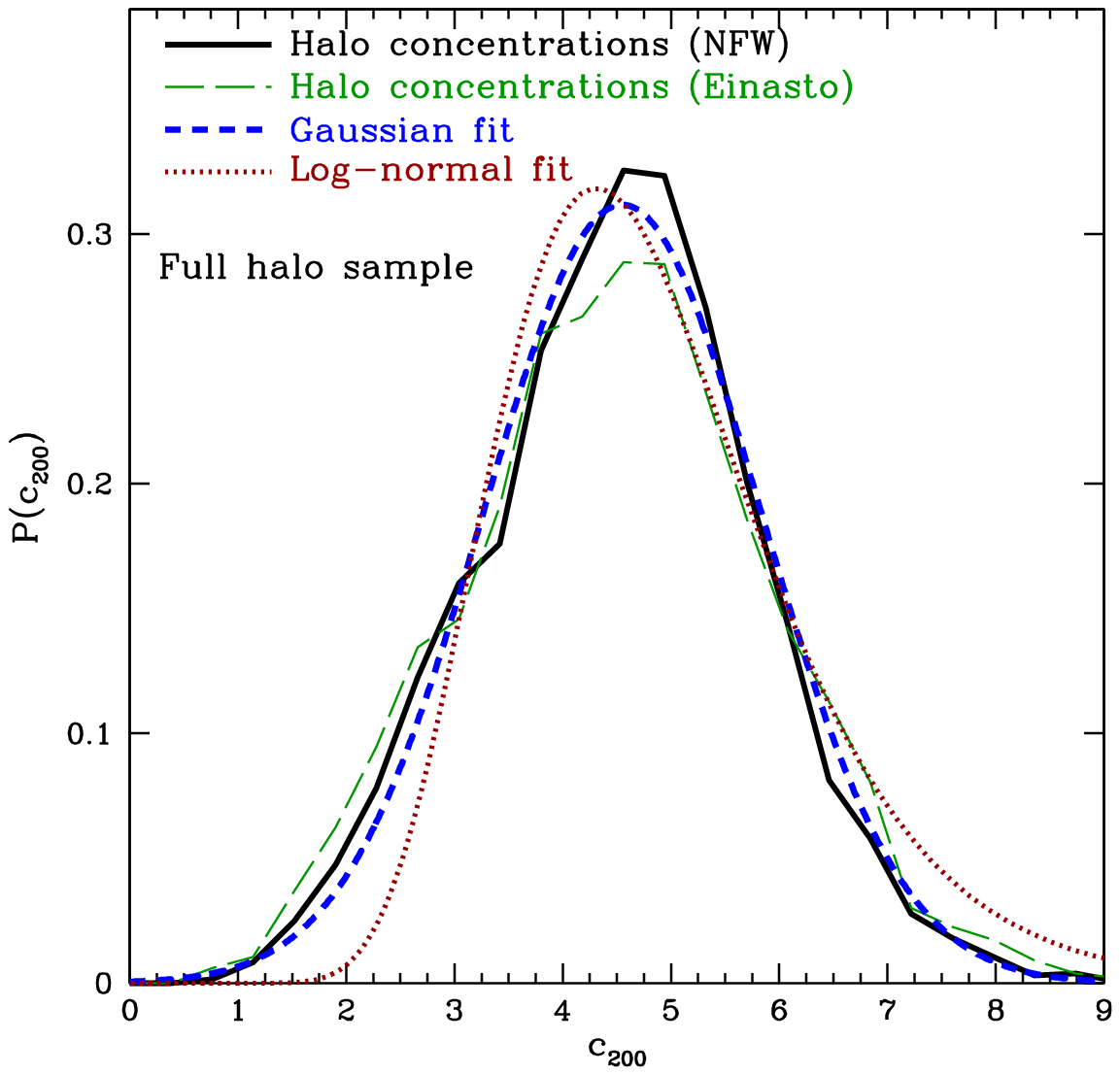} &
  \includegraphics[width=.5\textwidth]{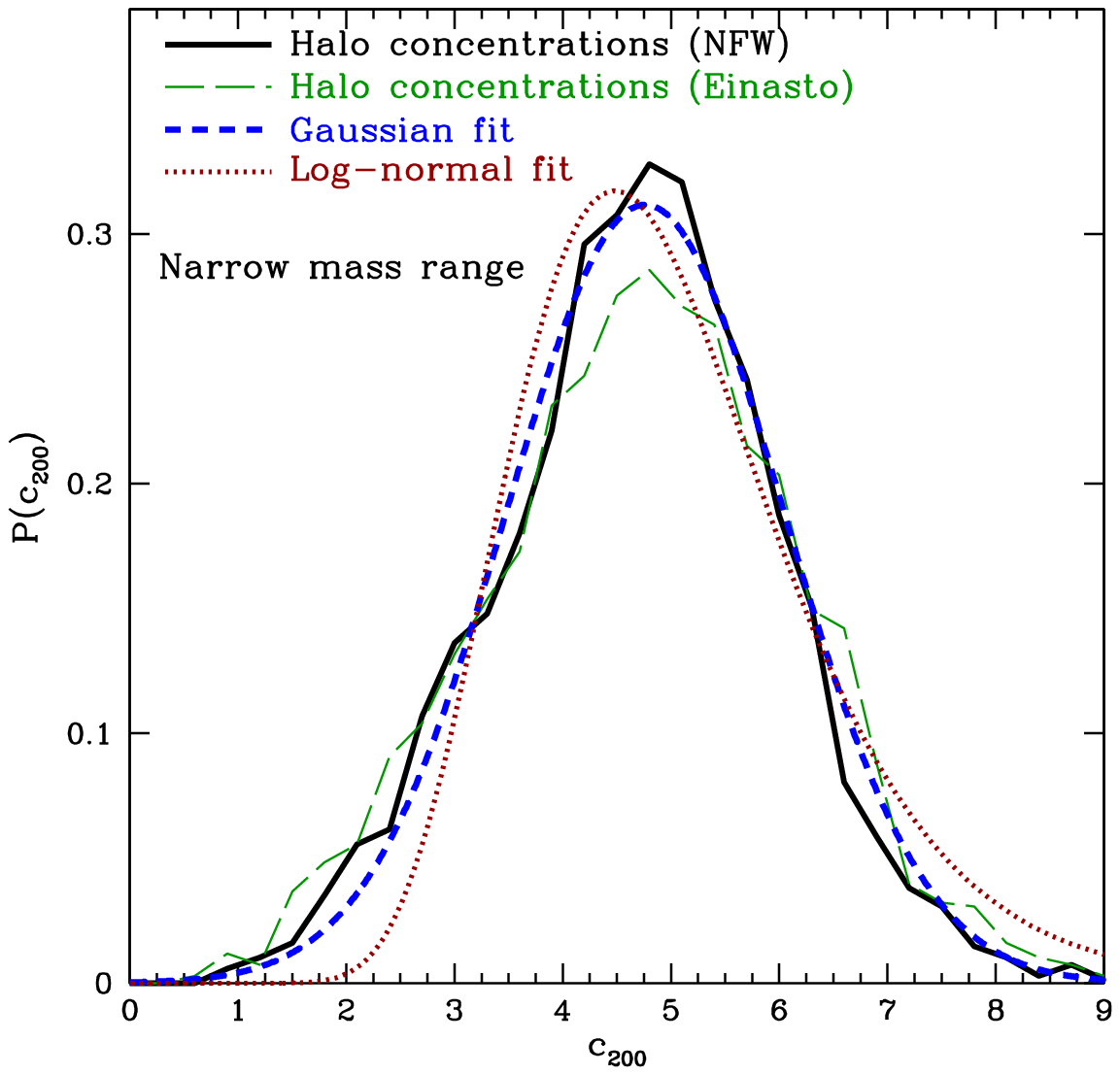} \\
\end{tabular}
  \caption{ {\it Left panel:} The probability distribution function
    (PDF) of concentrations for the 3,501 halos with $M \ge
    8.6\times10^{13} h^{-1}M_{\odot}$ shown against best-fit Gaussian
    with a mean $\langle c \rangle=4.55$, and standard deviation
    $\sigma_c=1.28$. This provides a better match than the best-fit
    log-normal distribution with a mean $\langle \log_{10}c \rangle =
    0.669$ and a standard deviation in $\log c$ of $\sigma_{\log
      c}=0.121$.  Concentration fits made to the NFW profile produce a
    modestly narrower distribution, even though the Einasto profile is
    overall a better description of the halo mass distribution.  {\it
      Right panel:} The distribution of concentrations in a narrow
    virial mass range of $8.6\times10^{13} \le [M/h^{-1}M_{\odot}] \le
    1.7\times10^{14}$ (2,276 halos). The shape of the concentration
    PDF remains Gaussian. }
  \label{concscatter}
\end{figure*}

\section{Simulations and halo catalogue}
\label{sec:thesimulation}

We utilize the gravity-only N-body particle {\it Millennium
  Simulation} of \cite{mill}.  This simulation evolves 2160$^{3}$
particles in a periodic box of 500$h^{-1}$Mpc using the gravity solver
code {\sc lgadget2} (\citealt{mill}), a modified version of the
publicly available {\sc gadget2} (\citealt{gadget2}).  Particle mass
is $8.6\times10^{8} h^{-1} M_{\odot}$. The cosmological parameters
used are $\Omega_m=0.25$, $\Omega_{\Lambda}=0.75$, hubble constant
$h=0.73$, $\Omega_b=0.045$, $n=1$, with power spectrum normalization
$\sigma_8=0.9$.  The matter power spectrum used to create initial
conditions is produced using {\small CMBFAST} (\citealt{cmbfast}).

For this study, it is important to use as large a statistical sample
of halos as possible. In addition, these halos must be resolved in the
innermost regions in which we are interested.  We thus consider all
halos with more than 10$^5$ particles, corresponding to $M_{vir} \geq
8.6\times10^{13} h^{-1} M_{\odot}$.  This results in 3,501 halos. The
density profiles for these halos are resolved down to $\sim
1-2\%r_{vir}$, based on convergence tests in \cite{mooreprores},
\cite{powerres}, and \cite{reedpro} for halos with similar numbers of
particles.

We define halos and their centers using the same procedure as was done
in \cite{netoconcs} and \cite{gaoconcs}.  Halos are identified
initially using {\it friends-of-friends} with linking length of 0.2
times the mean inter-particle spacing.  Halos are centered on the
location of the deepest potential of the main subhalo.  Halo mass is
then determined by a sphere of 95.4 times critical density
($M_{vir}$), and additionally by a sphere of 200 times critical
density ($M_{200}$).  We determine a spherically-averaged
logarithmically binned density profile for each halo.

\section{Cosmological Variations in the Halo Profile}
\label{sec:resultsprofile}

In this section we discuss the cosmological variance in the properties
that describe the profile of dark matter halos.

\subsection{Halo Concentrations}
\label{subsec:concs}

Halo concentrations have been shown to have significant halo-to-halo
scatter, with a median that decreases with increasing mass and
redshift (\eg \citealt{bullock}; \citealt{netoconcs};
\citealt{gaoconcs}; \citealt{duffy}; \citealt{maccioconcs}).  Early
work suggested that the distribution of halo concentrations is
log-normal (\citealt{jing}; \citealt{bullock}).  However, larger
samples of higher resolution halos reveal significant departures from
a log-normal scatter, primarily due to a tail of low concentrations
inferred from ``unrelaxed'' halos (\citealt{netoconcs};
\citealt{maccioconcs}), which tend to conform poorly to smooth
functional fits (\citealt{lukic09}).  In fact, the distribution of
halo concentrations is very well described by a simple Gaussian, as
noted by \cite{lukic09}, when all (relaxed and unrelaxed) halos are
considered.  In Fig.~\ref{concscatter} we show the concentration
probability distribution function (PDF) from our full sample.  We find
that a Gaussian description of concentrations is a better fit than a
log-normal distribution to the halos in our sample. More complicated
functional descriptions of halo concentrations such as that suggested
by \cite{netoconcs} or by \cite{maccioconcs} appear unnecessary to
describe our data (which consists of the high mass halo subset of the
halos of \citealt{netoconcs} and \citealt{gaoconcs}).

\begin{figure*}
\centering
\begin{tabular}{cc}
  \includegraphics[width=.5\textwidth]{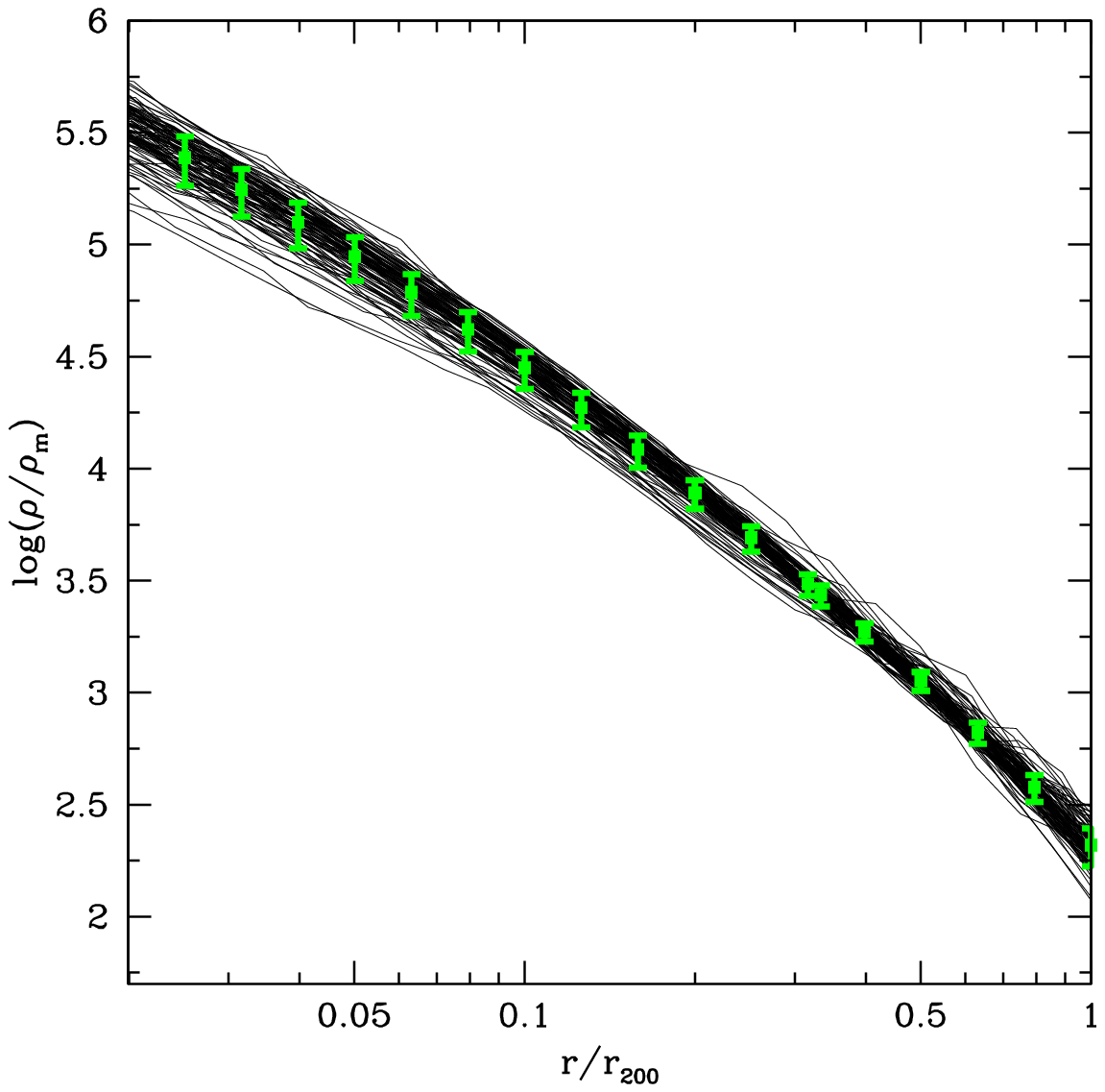} &
  \includegraphics[width=.5\textwidth]{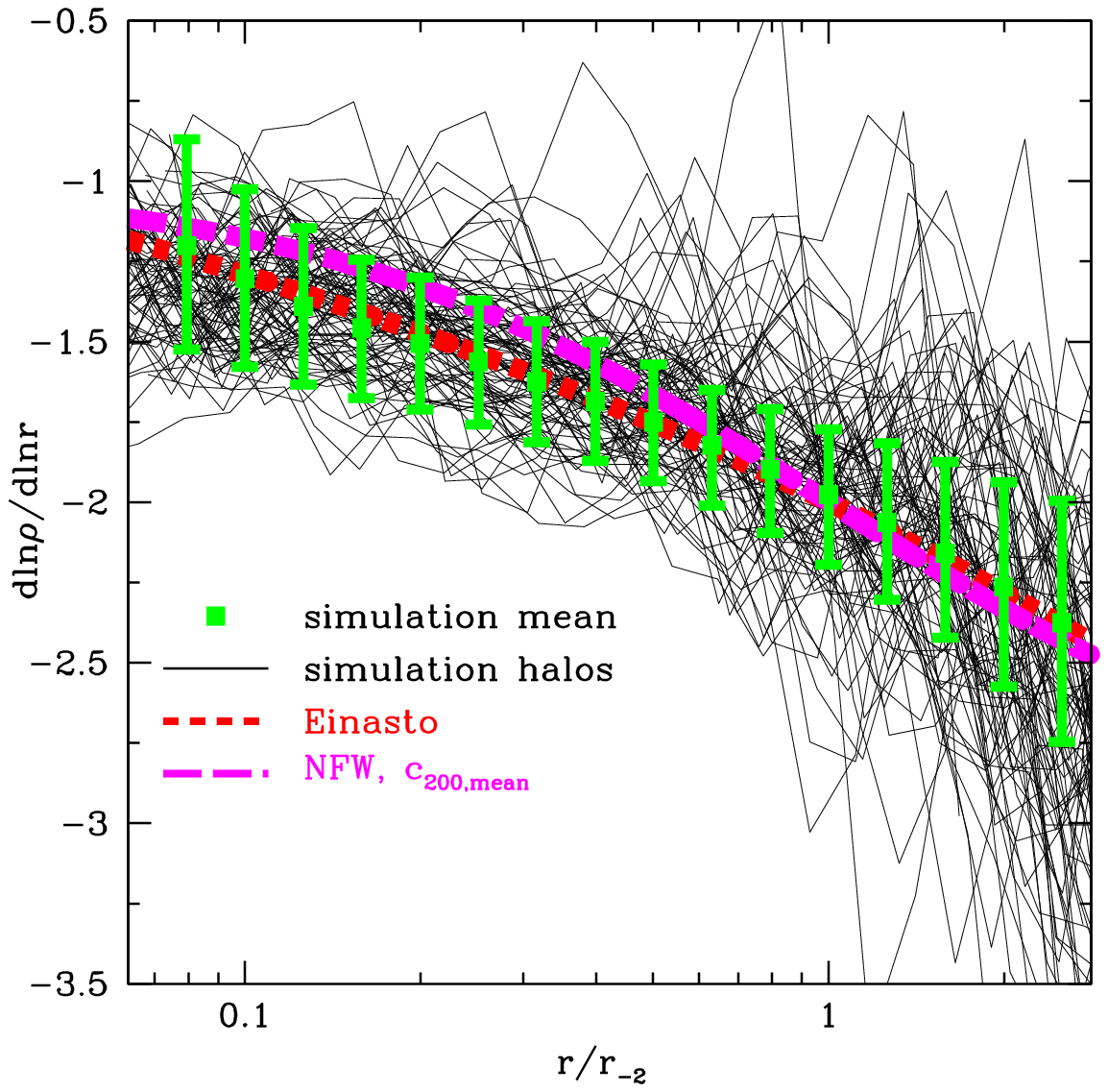}\\
\end{tabular}
  \caption{ {\it Left panel}: Density profiles for 100 random halos
    from the sample.  Points with error bars denote the mean and the
    $1 \sigma$ scatter of the distribution for the full halo sample.
    {\it Right panel}: Logarithmic density slope for the 100 largest
    halos.  Slopes are normalized to the scale radius to remove the
    effect of scatter in halo concentrations.  An Einasto profile with
    $\alpha=0.19$ is a good match to the mean simulation slope.  }
  \label{rhosloscatter}
\end{figure*}

In estimating halo concentrations, we consider both the Einasto
profile (Eq.~\ref{eqnein}-\ref{eq:normalization}) and the NFW profile
(NFW 1996; 1997):
\begin{equation}
\rho={4\rho_{-2} \over (r/r_{-2})(1+r/r_{-2})^2}.
\label{eqnnfw} 
\end{equation}
A concentration defined by the Einasto profile is, in principle,
equivalent to that defined by the NFW profile, both having a scale
radius at $r_{-2}$.  However, because simulation halos tend to better
match the Einasto form, which is steeper than NFW at the smallest
radii, a concentration inferred from the NFW form can be biased high
or low, depending on the range of radii used in fitting (see
\citealt{gaoconcs}).  We find that although the Einasto profile
produces a better fit to stacked halos, the distribution of
concentrations is modestly narrower when fit according to an NFW
profile (see Fig. \ref{concscatter}), with approximately the same mean
value ($c_{200,NFW}=4.55$ versus $c_{200,Einasto}=4.50$).  The Einasto
distribution remains wider, whether or not the Einasto parameter
$\alpha$ is fixed or allowed to float as a free parameter in the fit.
For this reason (and also for convenience), we determine halo
concentrations by fitting the NFW profile (Eq. \ref{eqnnfw}) in the
remainder of this paper.  We stress that the shape of the Einasto-fit
distribution of concentrations is nearly identical to that of the
NFW-fit concentration PDF; the only difference is that the Einasto-fit
concentration PDF is slightly wider ($\sim 10\%$) Gaussian.  Profiles
are fit to logarithmic radial bins over a range of $0.05-1r_{200}$
with normalization set by the mass contained within $r_{200}$.

We have confirmed also that fixing the density normalization instead
of allowing it to float as a fit parameter has no significant effect
upon the derived concentrations.  As a further test, we show that the
relatively wide range in halo masses for our full sample does not
affect the shape of the distribution of concentrations.  With a
narrower sample mass range of $2\times$ in mass ($8.6\times10^{13} \le
[M_{vir}/h^{-1}M_\odot] \le 1.7\times10^{14}$), a Gaussian
concentration distribution is still preferred over log-normal (right
panel of Fig. \ref{concscatter}).  The similar shape for the narrower
mass range is not surprising because the mass dependence on mean halo
concentration is relatively weak; and more importantly, this shows
that the shape of the concentration distribution is not sensitive to
our choice of the width of the mass range for the halo sample used
throughout the paper.


\subsection{Halo Densities}
\label{subsec:density}

It is important to note that the Einasto profile has been found to fit
halos well for a ``stacked'' ensemble (\eg \citealt{hayashiwhite};
\citealt{gaoconcs}). However, the presence of substructure and other
peculiarities implies that any particular halo profile tends to have
significant variations from this mean smooth function.

In the left panel of Fig. ~\ref{rhosloscatter}, we show this scatter
in the density profile for 100 random halos.  The spread in densities
due to different concentrations is also apparent.  The right panel of
Fig.~\ref{rhosloscatter} shows that there is very large scatter in the
logarithmic slope of the density profile, plotted here versus
$r/r_{-2}$.  This representation removes differences that arise due to
concentration.  The increased scatter in density slopes at outer radii
is independent of mass within our sample (\ie similar behavior is seen
for the 100 least massive halos in the sample); large radii scatter is
likely enhanced by large substructures.  The mean slope of the
complete halo sample is well described by an Einasto profile, and is
poorly matched by an NFW profile.

\begin{figure}
  \includegraphics[width=.5\textwidth]{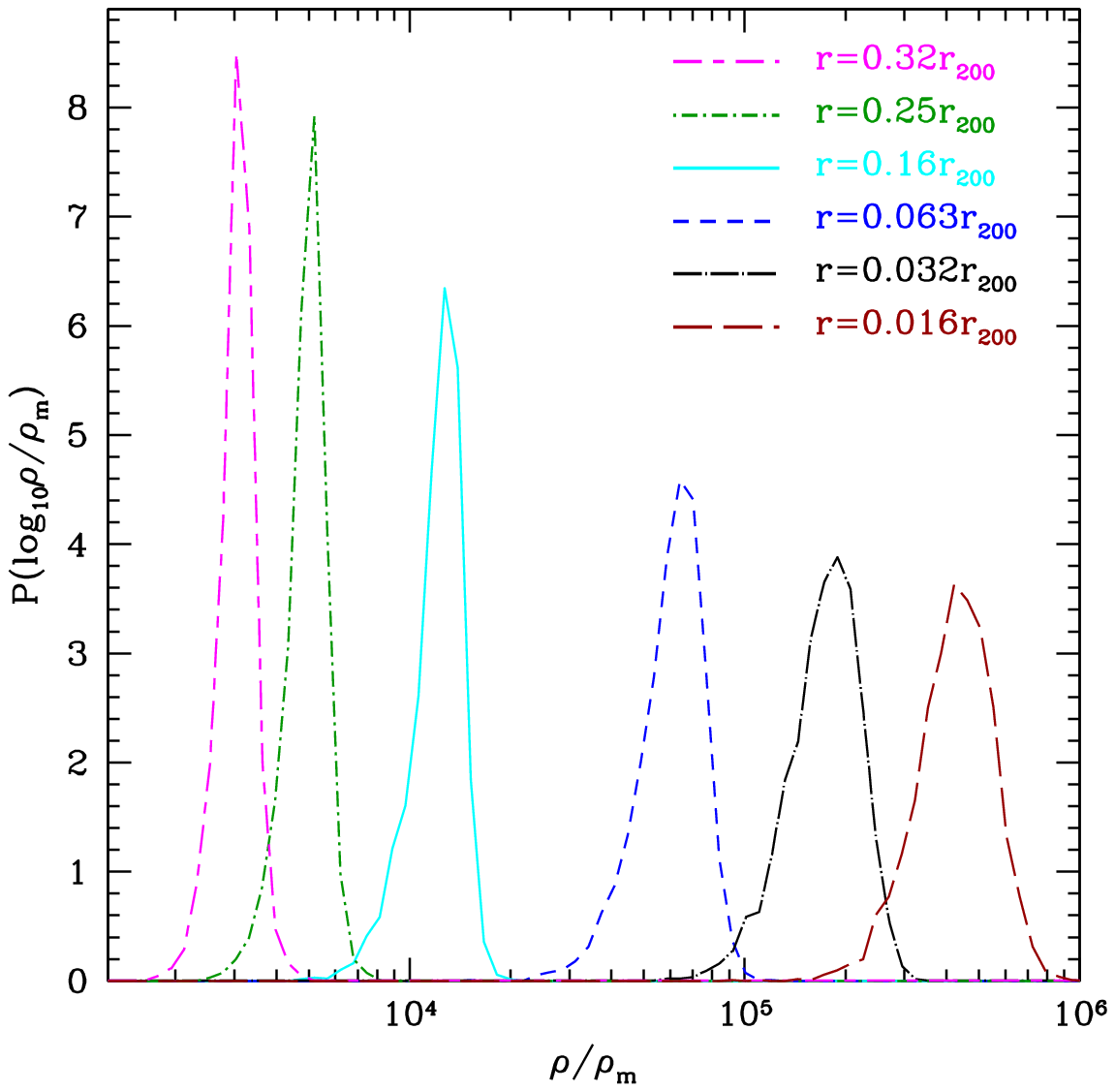}
  \caption{Halo-to-halo distribution in density at various radii
    relative to $r_{200}$ in the full halo sample.
Densities are plotted in units of mean density of the universe, $\rho_{\rm m}$.  
}
  \label{rhoscatterlog}
 \end{figure}

In Fig.~\ref{rhoscatterlog}, we show the probability distribution
function (PDF) of densities at various radii in spherically-averaged
shells from the halo sample.  The width of the halo-to-halo scatter of
density decreases toward larger radii.  A possible explanation for
this radial trend results from the fact that the central structure of
the halo is assembled at higher redshifts than the outer parts of the
halo (see \eg \citealt{fukupro}; \citealt{reedpro}).  If one assumes
that halo density at a particular radius correlates with the mean
density of the universe at the time of mass infall, then scatter in
mass assembly redshift from halo to halo would yield density
variations that would be larger nearer the center due to the $(1+z)^3$
evolution of the mean matter density.  Note that, at all radii, the
width of the distribution is small compared to the statistical
measurement uncertainty, which is estimated from Poisson counting of
particles in radial bins.

\begin{figure}
\centering
  \includegraphics[width=.5\textwidth]{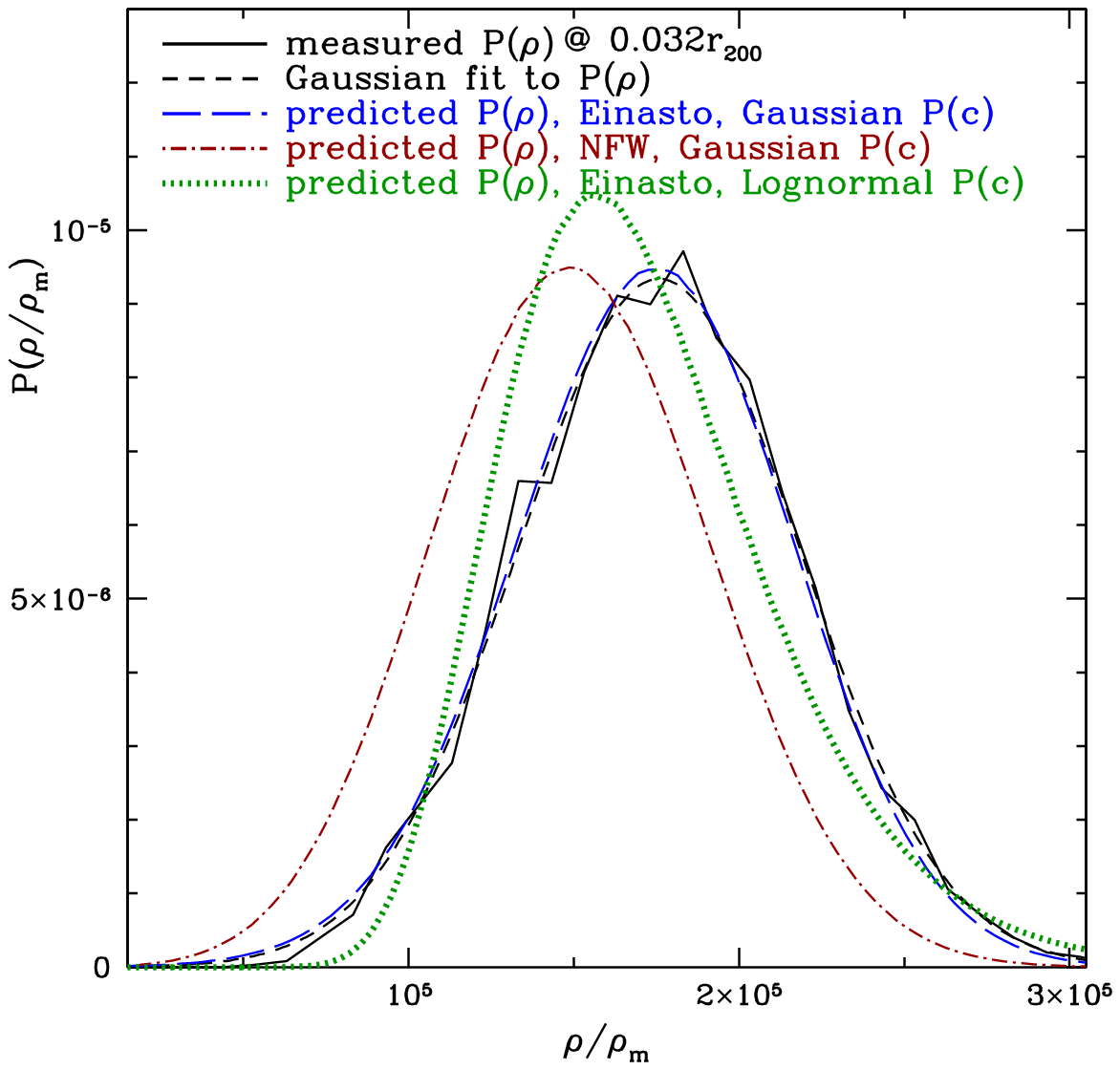} 
  \caption{ Measured density distribution at 0.032r$_{200}$, as an
    example, versus density distribution predicted by a deterministic
    Einasto profile and a Gaussian concentration distribution of width
    $\sigma_{c200}=1.28$, centered on $c_{200}=4.55$ to match the
    measured values of our sample.  Log-normal concentration
    prediction assumes the best-fit to our sample of
    $\sigma_{log10c}=0.121$ with median $c=4.66$.  Gaussian
    concentrations match the shape and peak position of the density
    distribution better than log-normal concentrations.  Note that the
    concentration distribution for all curves comes from an NFW
    concentration fit (even when applied to the Einasto profile),
    which produces essentially equivalent concentrations to an
    Einasto-fit (see Fig. \ref{concscatter} and
    \S~\ref{subsec:concs}); we use such NFW-derived concentrations
    here and hereafter.  }
\label{rhoscatterex}
\end{figure}

The halo density PDF is well-described by a Gaussian at each
radius. As an example, in Fig. \ref{rhoscatterex}, we show the density
distribution function at a radius of $r = 0.03r_{200}$.  This scatter
in densities is primarily due to the distribution in halo
concentrations rather than intra-halo departures from a smooth
functional form (\ie ``bumps'').  For small radii ($r\simlt r_{-2}$),
the Gaussian PDF of halo spherical shell densities is well-matched by
assuming that each halo is described by a deterministic Einasto radial
density profile whose concentration is drawn from a Gaussian
distribution (see Fig. \ref{rhoscatterex}).  This implies that the
distribution of halo concentrations fully determine the PDF of the
smooth density component, and provides additional support that the
form of the concentration PDF is Gaussian.  More specifically, the
effect of radial density ``bumps'' within the halo cannot be larger
than the effect of measurement uncertainty of individual halo
concentrations.  A log-normal concentration distribution is unable to
reproduce the density distribution shape or peak.  Unsurprisingly, an
assumption of an NFW profile does not match the density peak, although
it is able to produce the Gaussian shape and width.

\begin{figure}
\centering
  \includegraphics[width=.5\textwidth]{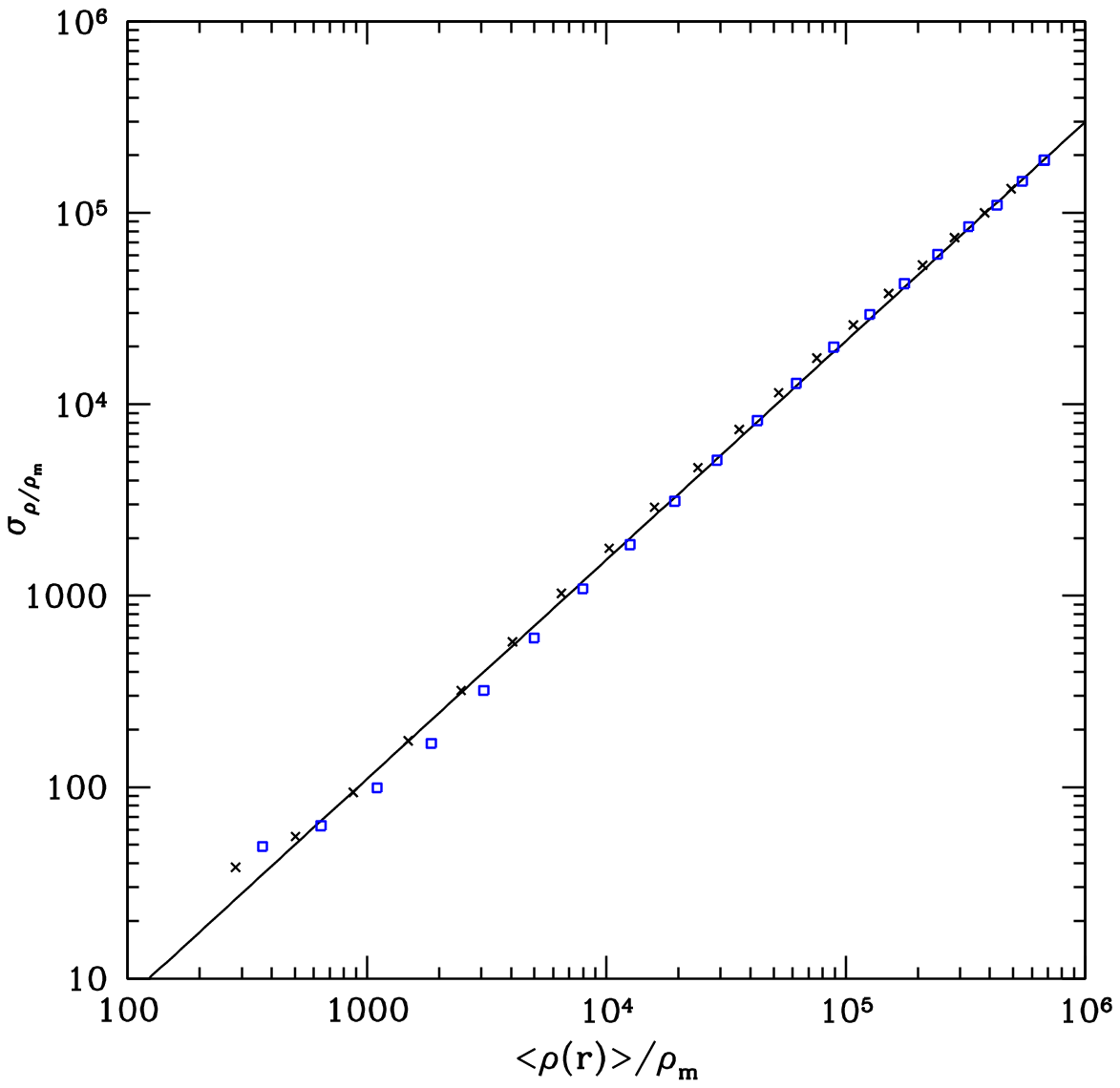} 
  \caption{ The width of the distribution of densities at various
    radii, $\sigma_\rho$, as determined by fitting a Gaussian, plotted
    as a function of mean density of our halo sample at each fitted radius
    for fixed $r/r_{vir}$ (black crosses) and for fixed $r/r_{200}$
    (blue squares).  The density distribution, $P(\rho)$, is
    well-described by a Gaussian distribution at all radii.  Line is a
    fit given by Eq. \ref{rhoscatterfit}.  }
\label{prhofit}
\end{figure}

At larger radii ($r\simgt r_{-2}$), the distribution remains Gaussian
but is wider than implied by the distribution of concentrations,
possibly due to increased scatter introduced from substructure or from
a lower degree of relaxation in the outskirts of these recently formed
cluster-size halos.  The distribution of densities in our data is
described to ($\sim 10\%$) accuracy by the following function:
\begin{equation}
{\log_{10}\sigma_{\rho} = 1.144 \, \log_{10} \langle \rho \rangle - 1.389},
\label{rhoscatterfit}
\end{equation}
where $ \langle \rho \rangle$ and $\sigma_{\rho}$ are found via a
Gaussian fit to the densities measured from the halo sample at a given
radius.  The agreement between the fit and the simulation data, shown
in Fig. \ref{prhofit}, over more than three orders of magnitude in
density is interesting.  However, we do not advocate that this is a
universal function; further work is required to determine whether this
relation between density and its scatter remains valid for lower halo
masses and different redshifts.  The moderate flattening of
$\sigma_{\rho}$ at low density reflects the relative widening of the
density PDF at large halo radii.

\begin{figure*}
\centering
\begin{tabular}{cc}
  \includegraphics[width=.5\textwidth]{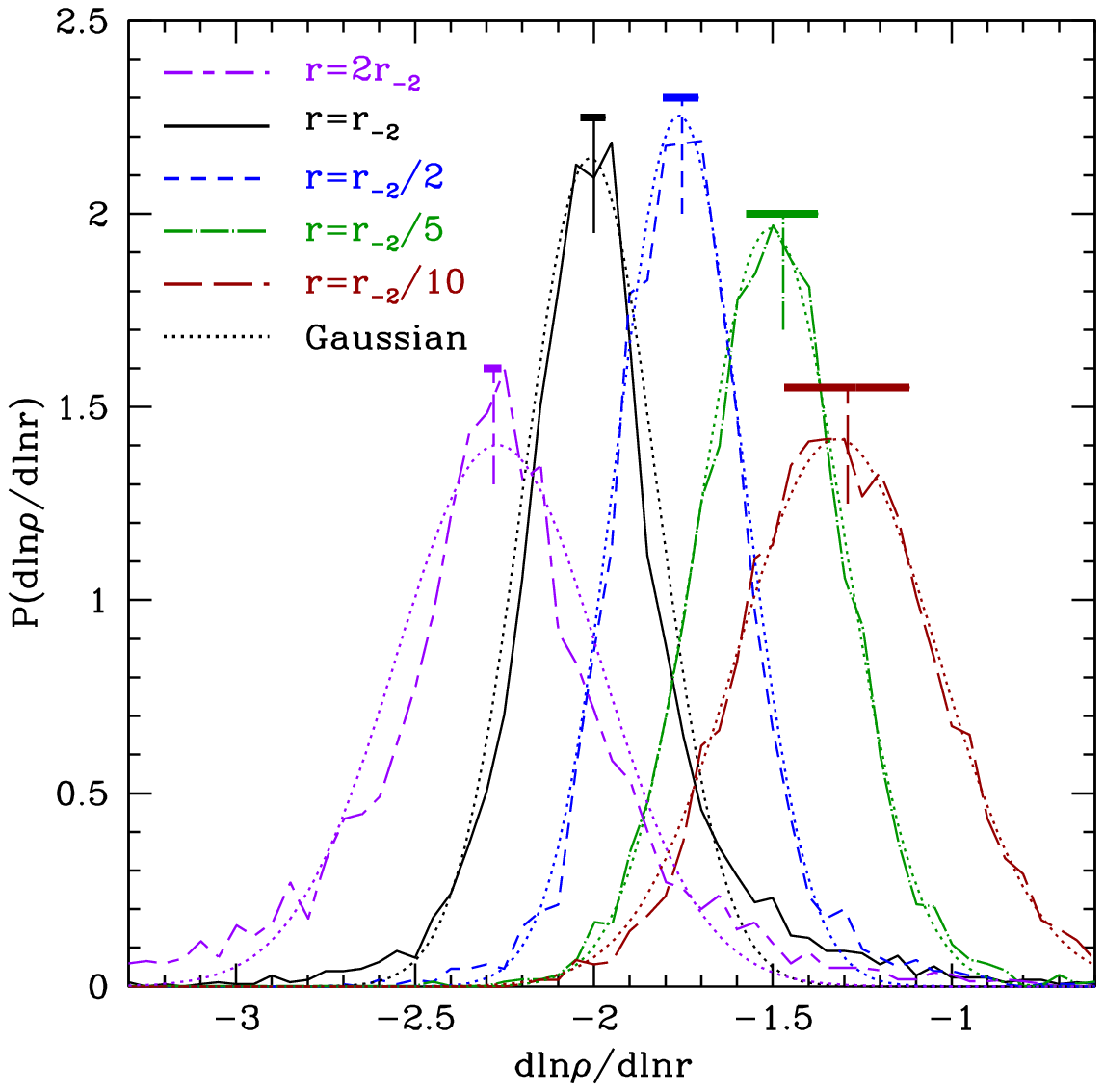} &
  \includegraphics[width=.5\textwidth]{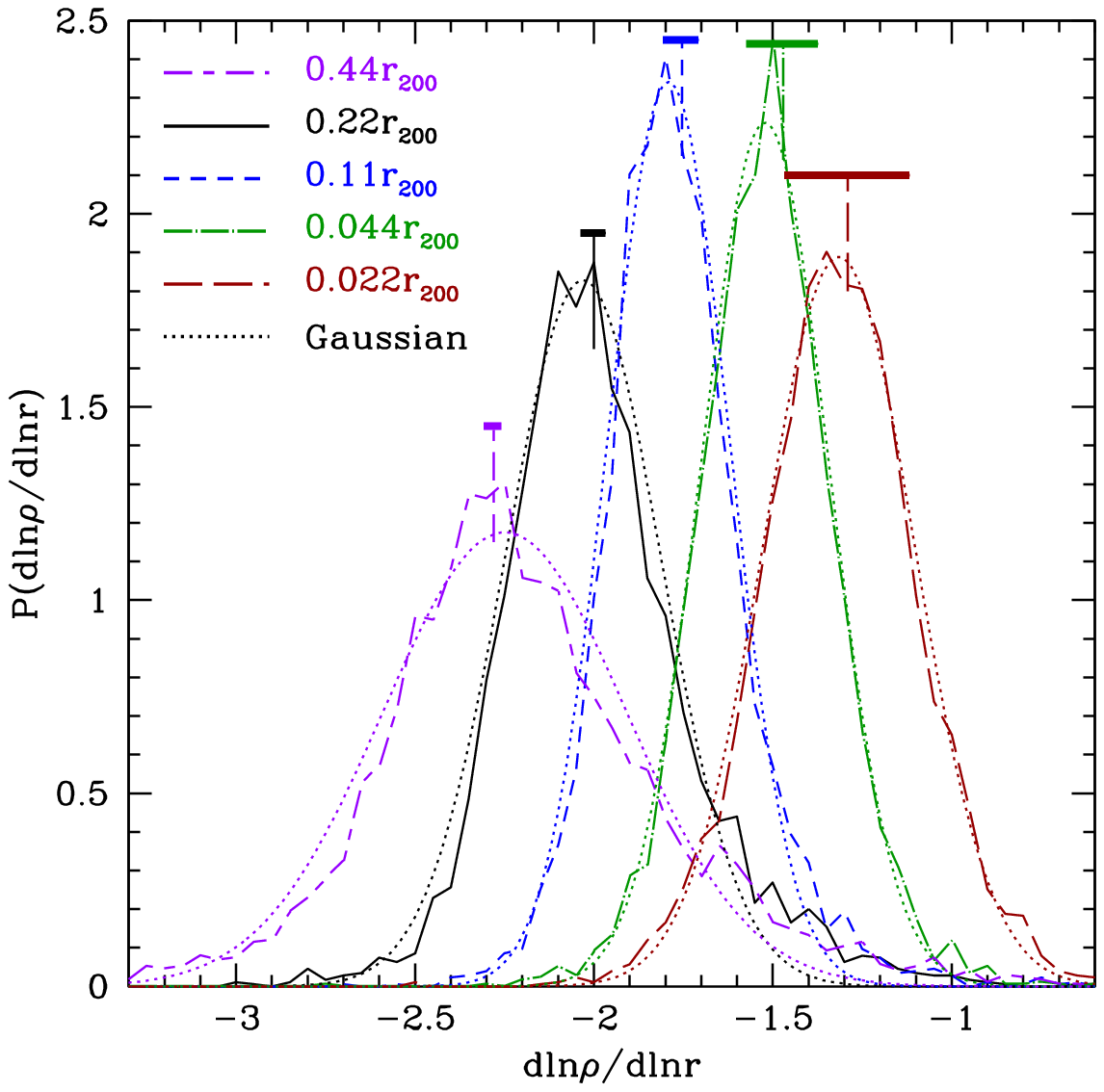}\\
\end{tabular}
  \caption{ Distribution in halo slopes at various radii (solid lines)
    relative to the scale radius ($r_{-2}=r_{200}/c_{200}$) (left),
    and relative to $r_{200}$ (right).  Dotted curves are best-fit
    Gaussians.  Vertical lines denote the slope of the Einasto profile
    with $\alpha=0.19$.  In the right panel, the Einasto profile
    slopes are shown for the mean halo concentration of 4.55.  Slopes
    are reasonably fit by a Gaussian with an average consistent with
    the Einasto profile and a width that is narrowest near $r_{-2}$.
    Again, these concentrations (\ie $r_{-2}$) are estimated from an
    NFW profile fit, which is essentially equivalent to a direct
    Einasto fit.  Horizontal lines indicate an estimate of the slope
    measurement uncertainty. }
  \label{scatter}
\end{figure*}

It is instructive to consider also the distribution of halo density
profile slopes.  In Fig. \ref{scatter}, we show the distribution of
the logarithmic radial slope of the density at various radii.  The PDF
of $d \ln \rho/d \ln r$ is well-described by a Gaussian distribution,
except perhaps at large radii where substructure or other effects
appear to result in wider than Gaussian tails.  The mean of this
distribution at each radius is consistent with the Einasto profiles
with $\alpha=0.19$.  Note that the Einasto profile with fixed $\alpha$
implies zero scatter in the PDF of halo slopes.  The radii in the
right panel of Fig. \ref{scatter} are chosen such that they should
have identical logarithmic slopes, assuming the Einasto profile, for
the mean halo concentration of 4.55.

For most radii, the width of the distribution of slopes is similar
whether measured in terms of $r_{-2}$ or in terms of $r_{200}$, apart
from the innermost plotted radius.  This is surprising because
differences in halo concentrations should contribute to the spread in
$d \ln \rho/d \ln r$ only at fixed $r/r_{200}$, and not at fixed
$r/r_{-2}$, according to the Einasto (or NFW) self-similar profile
form in which $d \ln \rho/d \ln (r/r_{-2})$ is independent of
concentration.  For this reason, we naively would have expected the
spread in density slopes to be smaller at fixed $r/r_{-2}$ than at
fixed $r/r_{200}$ (provided that $r_{-2}$ is determined accurately).
This suggests that intra-halo ``bumps'' rather than halo concentration
is the major contributor to scatter in the slope PDF.  In fact,
Eq. \ref{eqnein} implies that the distribution in density slopes at
fixed $r/r_{200}$ due to the concentration distribution should be
$\simlt 0.1$, which is much smaller than the actual spread which
ranges from $0.17 < \sigma_{d \ln \rho/d \ln r} < 0.34$ at the fixed
$r/r_{200}$ values shown.

At the smallest radius
of $r_{-2}/10$, the slope distribution width is likely dominated by
errors in the concentration measurement (left panel).  Presumably for
this reason, the PDF of the slopes is narrower when considered with
respect to $r_{200}$ at the smallest radii.  At each radius, we show
an estimate of the uncertainty in the slope measurement, based on
poisson noise from the average number of particles per radial bin.
The slope uncertainty is significantly smaller than the measured PDF,
except at the smallest plotted radius.  Because numerical problems are
most difficult to overcome at small radii, poisson uncertainty could
underestimate the true error.  We thus cannot rule out the possibility
that the broadened PDF at small radii may have numerical origins.
Indeed, increasing the minimum halo mass of the sample by a factor of
5 narrows the slope PDF at the smallest radii such that this effect is
significantly smaller.

\section{Effects of profile scatter on dark matter annihilation}
\label{sec:annil}


In this section we discuss the effect of the profile scatter on the
expected signal in $\gamma$-rays (or other byproduct) from dark matter
annihilation in halos.

In general, the total gamma-ray luminosity from a halo of mass $M$ is
given by the volume integral of the square of its mass distribution as
\begin{equation} 
{\cal L} = \frac{ \langle \sigma v \rangle}{M_\chi^2}\int_0^{R_v}
\rho_M^2(r)d^3 , 
\label{eq:luminosity}
\end{equation}
where $M_\chi$ is the mass of the dark matter particle, and $ \langle
\sigma v \rangle$ is the thermal average of the annihilation cross
section. Particle physics enters through the mass of the dark matter
particle, and through its total annihilation cross section. As an
example, in order to demonstrate the effects of density distributions
on the annihilation flux, we consider a dark matter particle with mass
$M_\chi = 400 {\rm GeV}$, with a total annihilation rate to $b
\bar{b}$ quarks given by $\langle \sigma v \rangle = 10^{-26} {\rm
  cm^3/s}$. We assume these values throughout the rest of this
manuscript, and note that in general, the assumed dark matter particle
properties affect the normalization of the results presented, and not
the shape of the distributions.

The ``dark luminosity'' of a halo is determined by the distribution of
its mass (see Eq.~\ref{eq:luminosity}).  We assume that the {\em mean}
distribution of dark matter in halos is described by the Einasto
profile, from Eq.\ref{eqnein}--\ref{eq:normalization}.  For the
remainder of this section, we focus on the dark luminosity of the
smooth component in a halo and thus we ignore the presence of
substructure and any associated ``boost'' that may contribute to the
annihilation luminosity as defined in Eq.~\ref{eq:luminosity}.

\subsection{The flux distribution as a function of mass}
\label{subsec:fluxmass}

\begin{figure}
\includegraphics[width=.5\textwidth]{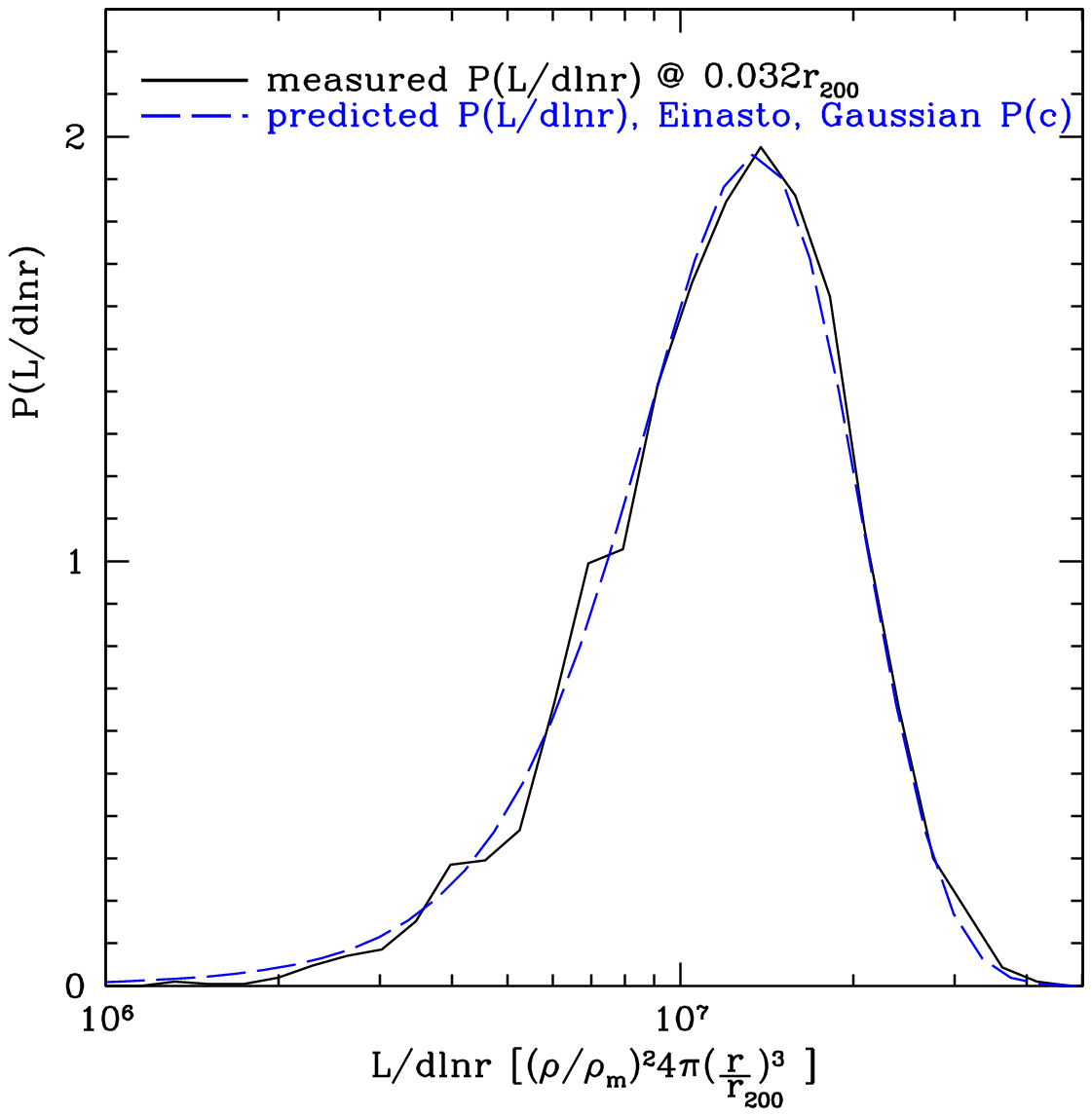}
  \caption{ Normalized dark matter annihilation ``luminosity'' per
    logarithmic radial interval ($4 \pi \rho^2 (r) r^3$) for the full
    halo sample.  The prediction assumes an Einasto profile with a
    Gaussian concentration scatter, as in Fig. \ref{rhoscatterex}.  }
\label{fig:rhosqscatter}
\end{figure}

We first confirm that we can capture the PDF of the local dark matter
annihilation volume emissivity (in spherical shells).  This is the
quantity that will be integrated to compute total halo annihilation
luminosity.  In Fig. \ref{fig:rhosqscatter}, we compare the measured
distribution of normalized differential annihilation luminosity per
logarithmic radial interval ($4 \pi \rho^2 (r) r^3$), shown here at
$r=0.032r_{200}$ as an example, with that from an Einasto profile
(with $\alpha = 0.19$) with the mean concentration and Gaussian
scatter of the halo sample (Eq.~\ref{eq:luminosity}).  We plot this
distribution in units of mean density and $r_{200}$ so that the
quantity is independent of halo mass.  The excellent agreement of
measured and predicted differential annihilation luminosity implies
that the concentration distribution with the assumption of an Einasto
profile is sufficient to estimate localized (in radius) dark matter
annihilation luminosity.  However, we have yet to establish that the
localized annihilation strength can be integrated to yield the correct
halo annihilation luminosity.  Correlations of density with radius
could result in large scatter in annihilation luminosity from halo to
halo.  For example, halos that happen to have enhanced density over
some extended range below $r_{-2}$, where the annihilation rate is
larger (albeit within a smaller volume), could have enhanced
annihilation luminosity versus halos with smoother profiles and
similar concentrations.  Although the results of
\S~\ref{sec:resultsprofile} imply that intra-halo radial correlations
are less important than the scatter in concentrations on local
density, the $\rho^2$-dependence of annihilation could lead to a
larger impact on halo annihilation rates.

\begin{figure*}
\centering
\begin{tabular}{cc}
  \includegraphics[width=.5\textwidth]{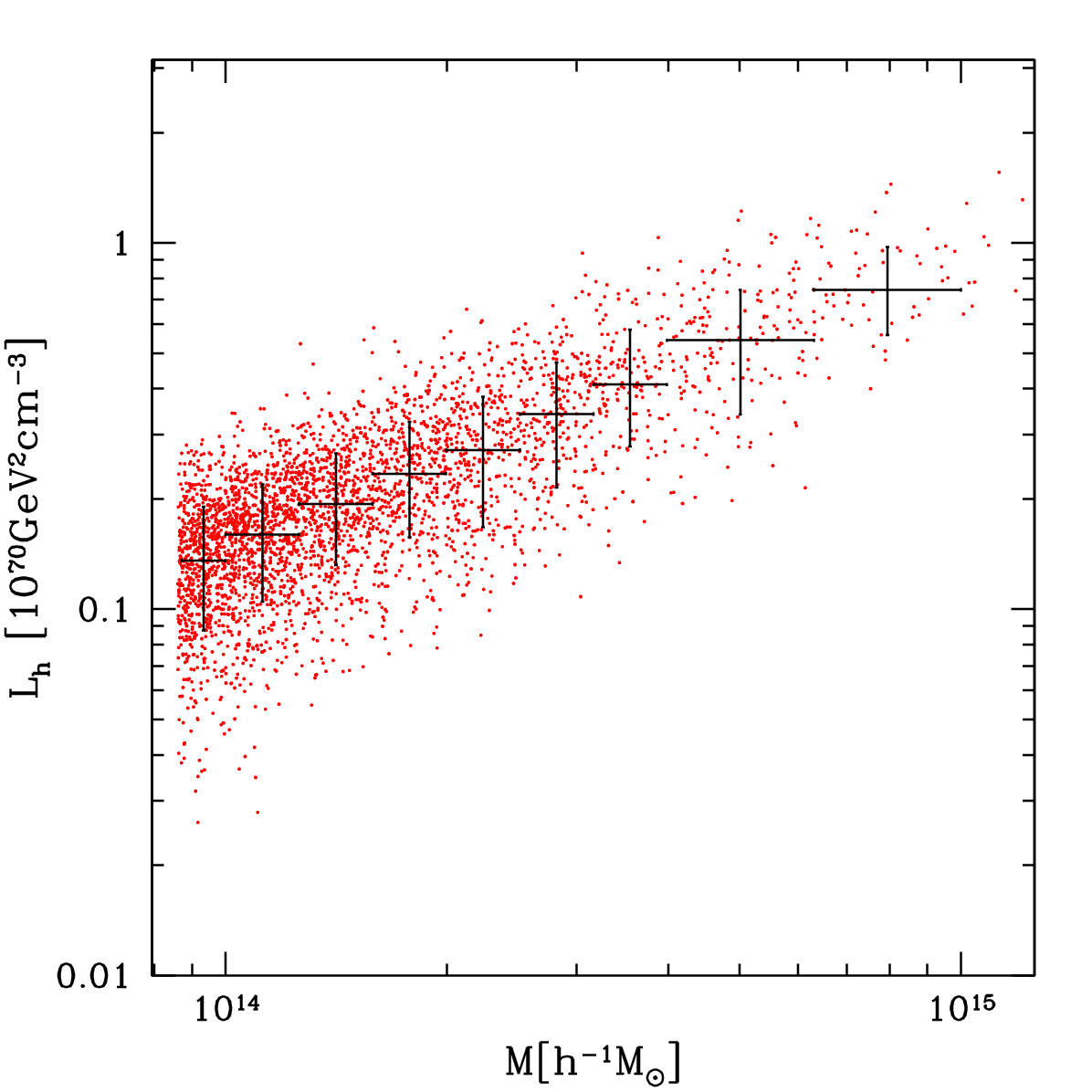} &
  \includegraphics[width=.5\textwidth]{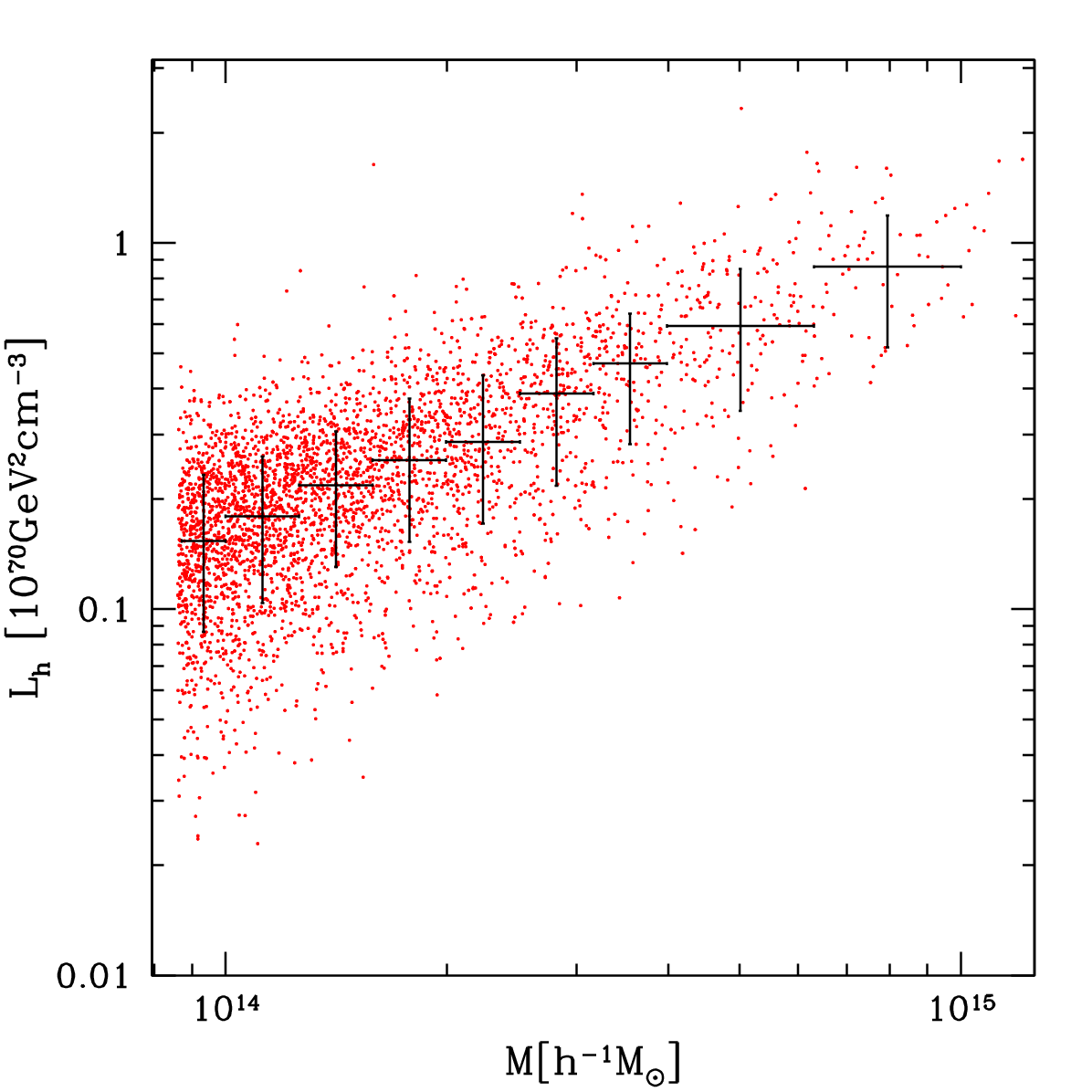} \\
\end{tabular}
  \caption{The inferred dark matter luminosities in units of $10^{70}
    {\mathrm {GeV}^2 \mathrm{cm}^{-3}}$. Each dot represents a dark
    matter halo, while the vertical bars represent the 68 percentile
    of the distribution in each virial mass bin, whose width is
    denoted by the horizontal bars.  {\it Left panel:} Luminosity
    computed directly from the density profile extracted from the
    simulation for each halo.  {\it Right panel:} Expected luminosity
    computed from fitting a concentration to each halo and assuming an
    Einasto density profile. }
  \label{fig:luminosities}
\end{figure*}

In order to assess the origin of the scatter in halo annihilation
luminosities and the contribution due to the scatter in
concentrations, we compute the dark luminosity of each simulation
halo, and compare with the expected luminosity given its measured
concentration.  We show in Fig.~\ref{fig:luminosities} the
distribution of halo dark luminosities computed directly from the
simulation halo density profile (left panel) together with the
prediction of the same quantity from an Einasto profile using the
individually measured concentration of each halo (right panel).  We
bin the data in mass, and determine the mean and 68 percentile of the
distribution.  The measured and predicted luminosities and 1 $\sigma$
scatter agree well; this suggests that {\it radial correlations in
  density should not prevent accurate estimation of the annihilation
  luminosity of a halo}.

Thus, the origin of the distribution of luminosities at each mass bin
is the distribution in concentrations, which correlate with formation
time, albeit with large scatter (see \eg \citealt{netoconcs}).  The
correlation between concentration and annihilation luminosity can be
modelled in the following manner for the smooth density component of
dark matter halos.  The normalization of the profile is proportional
to $\rho_{-2} \sim c_{200}^{\gamma}$, where $\gamma~\sim3$ for
$c_{200}\gg 1$, and $r_{-2} \sim c_{200}^{-1}$.  It follows that,
roughly speaking, the luminosity scales as $L \sim \rho_{-2}^2
r_{-2}^3\sim c_{200}^{2\gamma-3}$, leading to $L \sim c_{200}^{3}$ for
$c_{200}\gg 1$, although for concentrations typical of our clusters $L
\sim c_{200}^{1.5}$.  Note that it is difficult to distinguish between
the subtle differences between a Gaussian distribution of
concentrations and a log-normal one from Fig.~\ref{fig:luminosities}.
In addition, we note that the mean of the luminosity distribution at
each mass bin is roughly proportional to the mass of the halo. This is
to be expected as the luminosity of a dark matter distribution that is
described by a two-parameter profile (e.g., NFW, and/or Einasto) is $L
\sim \rho_{-2}^2 r_{-2}^3 \sim M^\beta$, where $\beta \approx 1$
because the dependence of concentration on mass is relatively weak.

\subsection{Cosmological $\gamma$-ray background}
\label{sec:background}

We now consider the contribution of the scatter in densities to the
cosmological $\gamma$-ray background: the annihilation flux integrated
over all halos at all masses and redshifts.  We are interested in the
effects of the non-universality of profiles to the expected gamma-ray
background.  In \S~\ref{subsec:density}, we showed for cluster halos
that the distribution in halo concentrations fully describes the
distribution in halo densities (at small radii).  This enables an
accurate estimate of the distribution in dark matter annihilation
luminosities (see \S~\ref{subsec:fluxmass}).  In order to estimate the
dark matter annihilation background, we assume that this holds for
halos of all masses at all redshifts.

We compute the gamma-ray background as
\begin{eqnarray} 
\frac{dN_\gamma}{dEdAdtd\Omega} &=& \frac{1}{4 \pi} \frac{c}{H_0}
\int_0^\infty \int_{M_{min}}^\infty \frac{\langle \sigma v
  \rangle}{M_\chi^2} \frac{dN_\gamma}{d[E(1+z)]} \\ &\times&
\frac{1}{h(z)}\frac{dn(z)}{dMdV} \int_0^{R_v} \rho_M^2(r)d^3r \, dM \,
dz, \nonumber
\end{eqnarray}
where $c$ is the speed of light, $H_0=0.7$ is the present value of the
Hubble constant, and $h(z) = \sqrt{ \Omega_M(1 + z)^3 +
  \Omega_\Lambda}$. We assume that $\Omega_M=0.27$ and
$\Omega_\Lambda=0.73$.  We use an Einasto density profile parameter of
$\alpha = 0.1645$, the value for the ``typical'' halo mass formed from
a $1\sigma$ peak in the mass-density field as given in \cite{gaoconcs}
(and we ignore any mass and redshift dependence of $\alpha$).  The
quantity $dN_\gamma/d[E(1+z)]$ is the spectrum of the emitted photons
at a source energy of $E(1+z)$, and the mass function of objects of
mass $M$ at redshift $z$ is $dn(z)/dMdV$. We use the mass function of
\cite{reedmf07}.

We assume that the annihilation proceeds to a $b\bar{b}$ quark final
state, and that the distribution in the number of gamma-rays emitted
per source energy interval $E_s$ are described by the functional form
given in \cite{bergstrom}, namely,
\begin{equation} 
\frac{dN}{dE_s} = \frac{1}{M_\chi}\left(\frac{E_s}{M_\chi}
\right)^{-3/2} {\rm exp} \left[ - 10.7 \frac{E_s}{M_\chi}\right] .
\end{equation}

We estimate the cosmological gamma-ray background for three different
halo concentration distributions. First, we assume an Einasto
functional form of the density as a function of radius and a
one-to-one dependence of concentrations on mass and redshift as given
in \cite{maccio}; this is case ``Einasto''.  However, halo
concentrations exhibit a distribution at fixed halo mass and redshift
(see \S~\ref{subsec:concs}).  As such, we consider a second case,
``Einasto $+$ Gaussian'', in which halo concentrations instead follow
a Gaussian distribution, given by $\sigma_G = 0.283 c_{200}$, where
$c_{200}$ remains the \cite{maccio} concentration as a function of
mass and redshift, and the Einasto profile still describes the density
distribution.  This value for the Gaussian width corresponds to the
fractional width found in our halo sample.  Finally, we consider case
``Einasto $+$ LogNormal'' where the concentration distribution is
log-normal about the mean concentration value, with dispersion given
by $\sigma_{LN} = 0.121$, and all other aspects of the background
calculation (\ie Einasto profile, mean concentration-mass-redshift
relation) remain the same as the two other cases.

\begin{figure}
  \includegraphics[width=.5\textwidth]{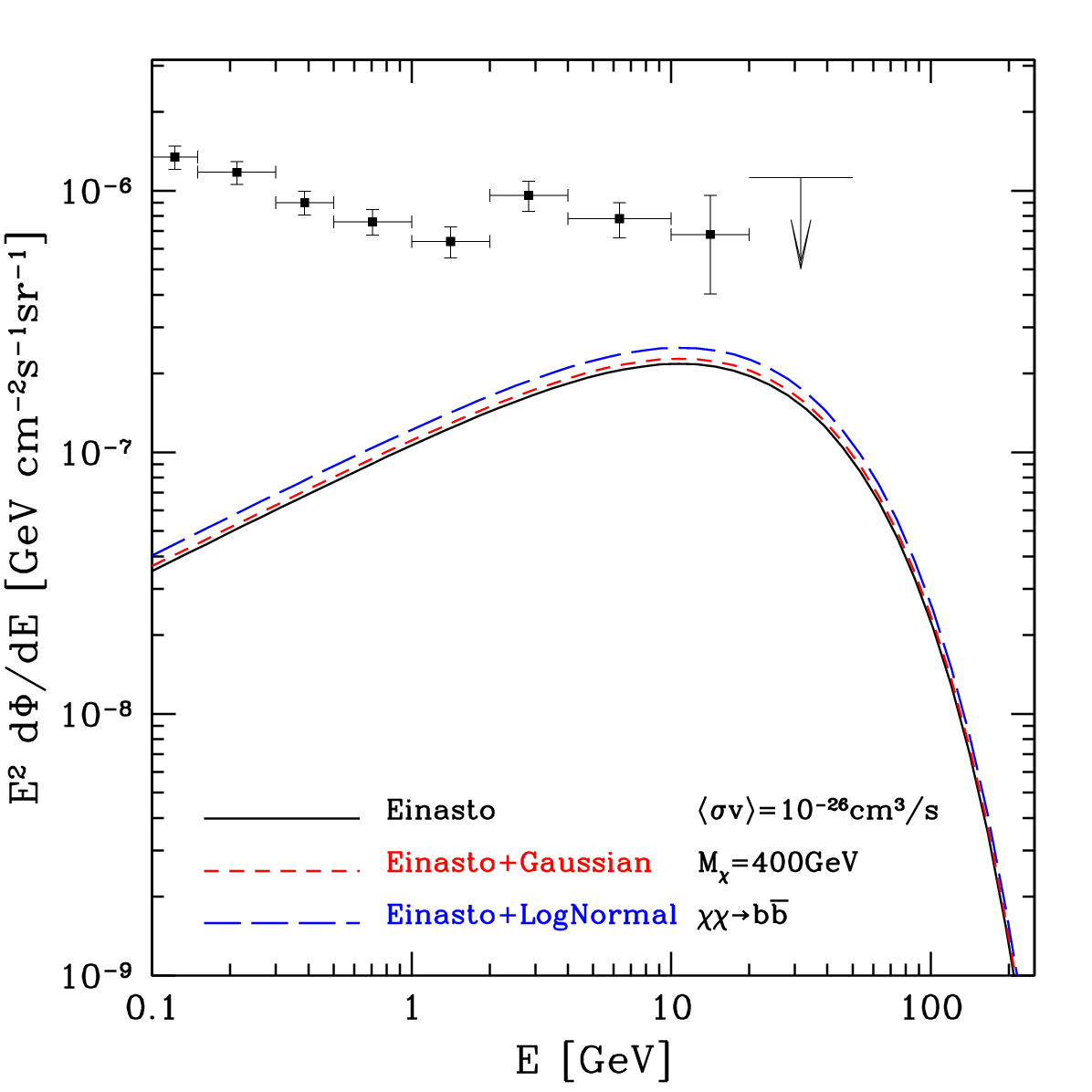}
  \caption{The effects of a Gaussian ({\em short-dashed}) and
    log-normal ({\em long-dashed}) distribution of concentration
    parameters to the cosmological extragalactic dark matter
    annihilation background. The {\em solid} curve represents the
    background computed by assigning a fixed $c(M)$ relationship
    without a spread. In all cases, the background is assumed to be
    due to a 400 GeV WIMP annihilating with a cross section of
    $10^{-26} {\mathrm cm^3 s^{-1}}$ into a $b \bar{b}$ quark pair.
    Data points are background measurements from the {\small EGRET}
    satellite (\citealt{strong}), which could include a contribution
    from dark matter annihilation.  }
  \label{fig:background}
\end{figure}

In Fig.~\ref{fig:background}, we show the expected cosmological
gamma-ray background for the three different distributions of dark
matter. As expected, in the presence of a spread in the distribution
of concentrations, the annihilation flux is increased relative to the
case where there is a one-to-one mapping between concentration and
mass.  We find that a log-normal distribution would give rise to
approximately a 10\% increase in the annihilation flux relative to our
preferred Gaussian distribution of concentrations (which is only a few
percent higher than the simple case of no distribution in
concentrations).

\begin{figure}
  \includegraphics[width=.5\textwidth]{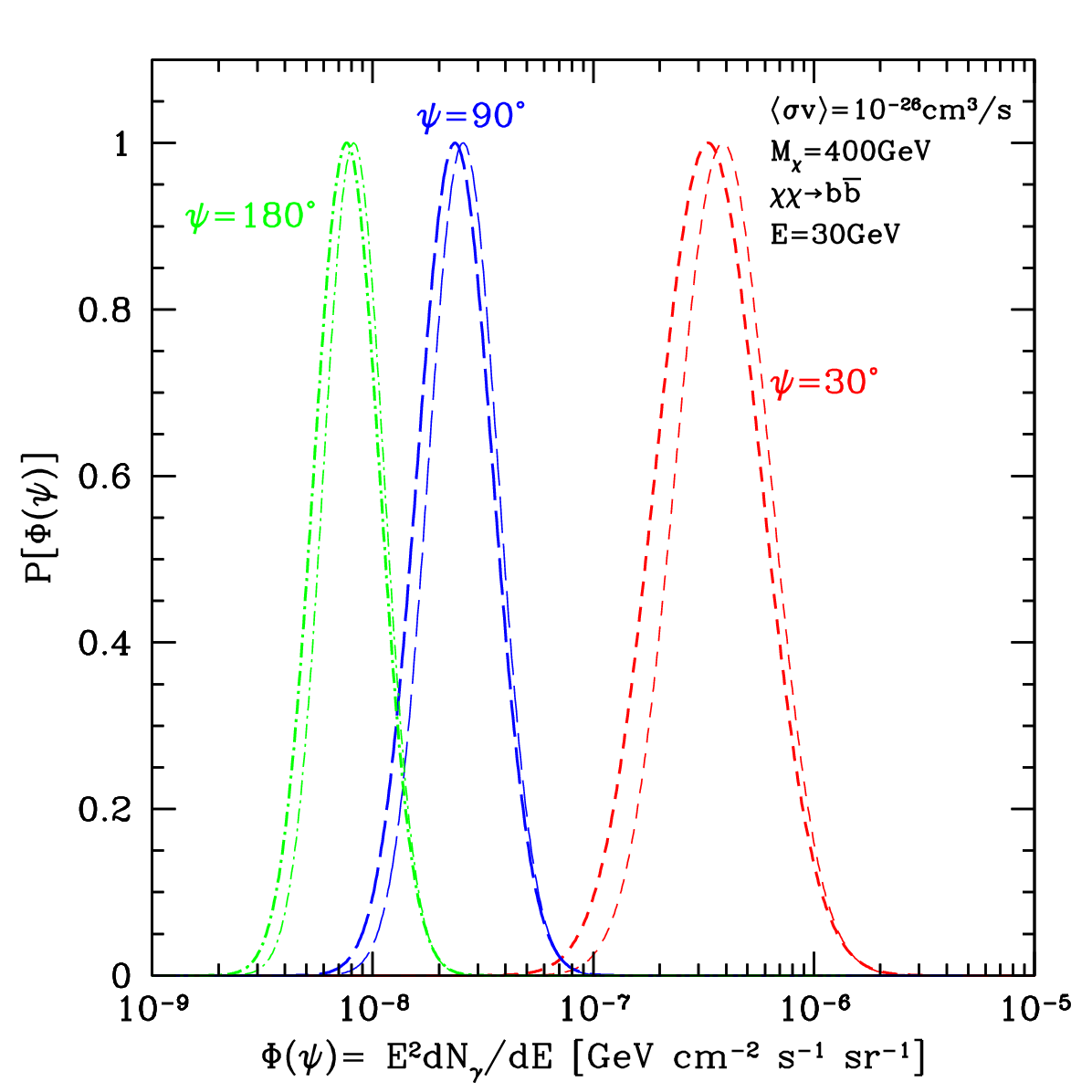}
  \caption{ The probability distribution function of gamma-ray flux
    along a line of sight at $30^\circ$ ({\em short-dashed}),
    $90^\circ$ ({\em long-dashed}) and $180^\circ$ ({\em dot-dashed})
    degrees from the Galactic center for a Gaussian concentration
    distribution ({\it thick curves}) and a log-normal concentration
    distribution ({\it thin curves}).  Note the smaller width of the
    distribution at high Galactic angles relative to the inner parts
    of the halo. Particle physics parameters are same as
    Fig.~\ref{fig:background}.  }
  \label{fig:MWPDFflux}
\end{figure}

\begin{figure}
  \includegraphics[width=.5\textwidth]{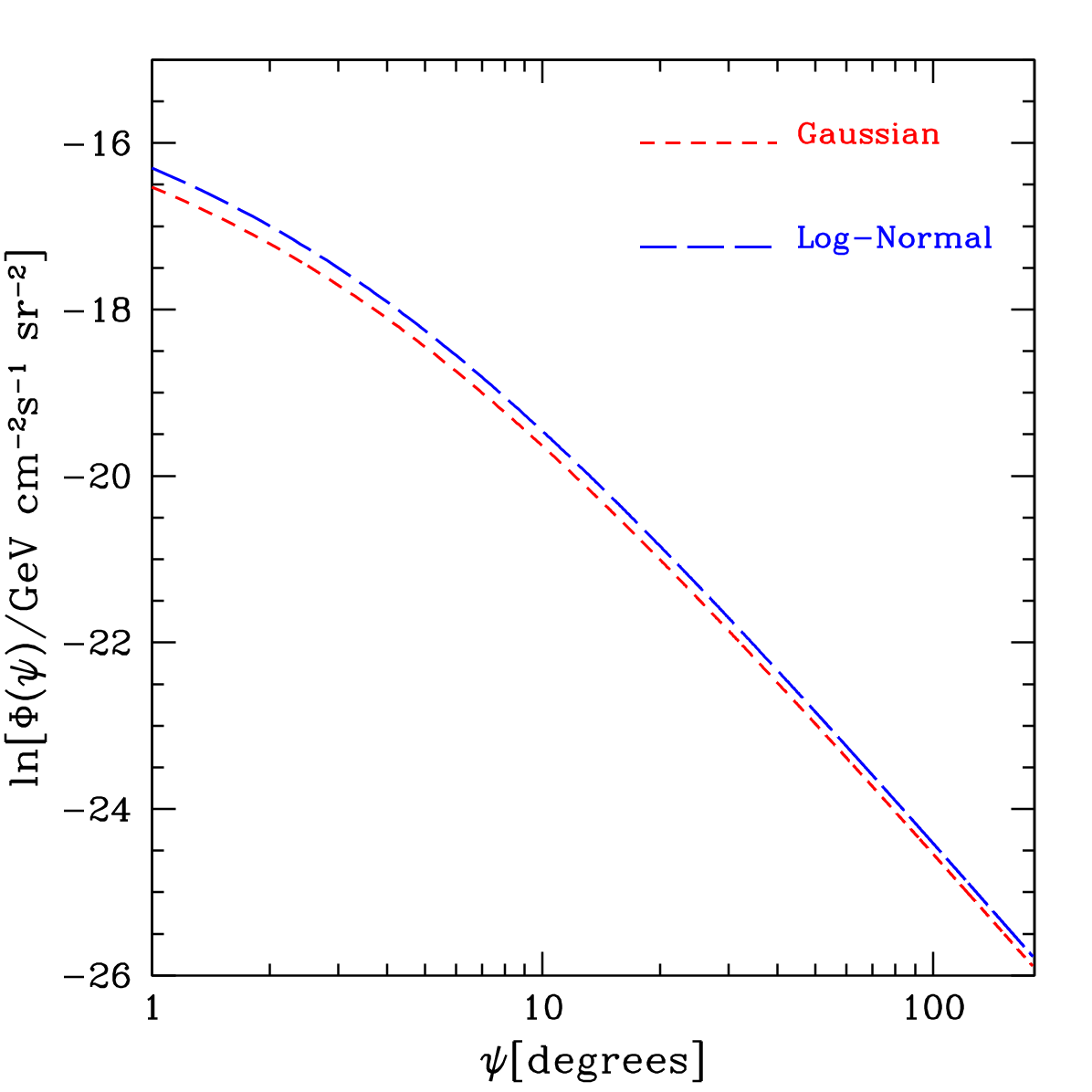}
  \caption{The dependence of the peak of the distributions
    (Gaussian/log-normal) to the angular distance of the line of sight
    with the Galactic center. Log-normal distribution in
    concentrations leads to a slightly higher peak in the flux
    probability distribution function. }
  \label{fig:MWFluxprop1}
\end{figure}

\begin{figure}
  \includegraphics[width=.5\textwidth]{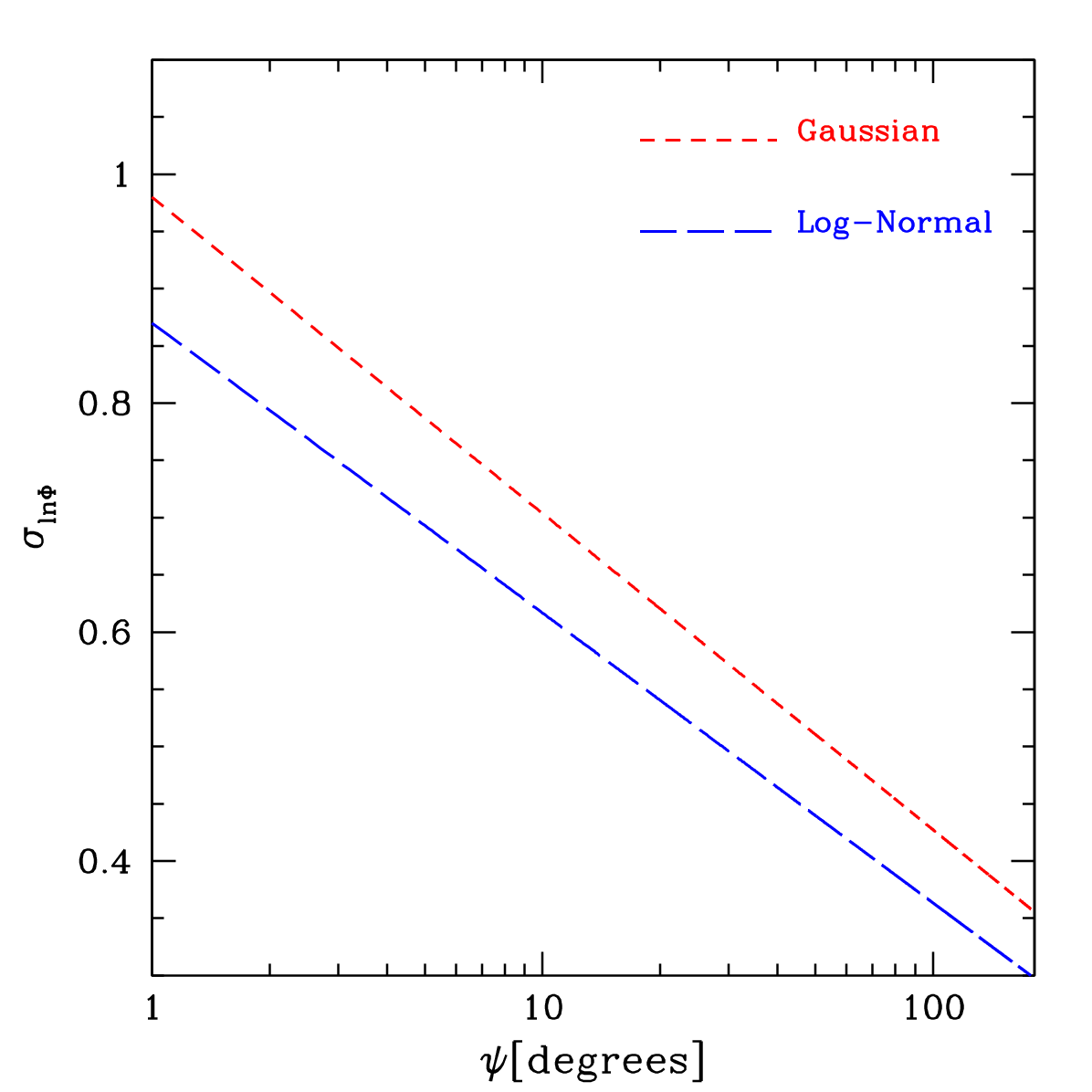}
  \caption{The dependence of the width of the distributions
    (Gaussian/log-normal) to the angular distance of the line of sight
    with the Galactic center. Log-normal distribution in
    concentrations leads to a slightly narrower flux probability
    distribution function.}
  \label{fig:MWFluxprop2}
\end{figure}

\subsection{The annihilation flux due to the smooth distribution of
  dark matter in the Milky Way} 
\label{sec:mwannil}

We now discuss the impact of the halo density probability distribution
function on the Milky Way annihilation flux along different lines of
sight.  The expected flux at a particular angle $\psi$ with respect to
the Galactic center depends on the distribution of dark matter
densities along that line of sight. As that is an outcome of the
particular concentration of the Galactic Halo, drawn from a
distribution of possible concentrations, there is a distribution of
expected fluxes for each line of sight.  Our calculations of the PDF
of dark matter annihilation within the Milky Way assume that the PDF
of the spherically-averaged Galactic dark matter density and
annihilation are well-described by an Einasto profile and the
corresponding PDF of halo concentrations, as we have shown to be the
case for cluster-massed halos.

The line of sight flux at an angle $\psi$ with respect to the Galactic
center can be written as
\begin{equation} 
\frac{dN_\gamma(\psi)}{dEdAdtd\Omega} = \frac{1}{4 \pi} \frac{\langle \sigma v
  \rangle}{M_\chi^2} \frac{dN_\gamma}{dE} \, \int_0^{\ell_{\rm max}} 
\rho^2_{MW}[r(\ell, \psi)] d \ell, 
\label{eq:LOS}
\end{equation}
where $\ell_{\rm max} = d_\odot [ \cos \psi + \sqrt{
    (R_{MW}/d_\odot)^2 - \sin^2\psi}] $ and $r = \sqrt{ d_\odot^2 +
  \ell^2 - 2 d_\odot \ell \cos \psi}$.  We take the distance of the
Sun from the Galactic center to be $d_\odot = 8.5 {\rm kpc}$
(consistent with \citealt{gillessen}), and the radius of the Milky Way
halo $R_{200,MW} = 250 {\rm kpc}$, which implies
$M_{200}=1.8\times10^{12}\msun$ and
$M_{vir}\simeq=2.\times10^{12}\msun$ (consistent with \eg
\citealt{guo}).  We assume an Einasto profile parameter $\alpha =
0.1645$, consistent with that for a halo of Milky Way mass
(\cite{gaoconcs}).

In Fig.~\ref{fig:MWPDFflux}, we show the expected flux distribution at
various angles with respect to the Galactic center computed using
Eq.~\ref{eq:LOS}. We calculate the flux distribution for two cases,
first where the distribution of concentrations is Gaussian, and second
where the distribution of concentrations follows a log-normal
distribution.  We find that for a log-normal distribution of
concentrations the width of the flux distribution along a line of
sight is slightly narrower, while at the same time, the mean of the
distribution is slightly higher. This is to be expected as the
annihilation rate is sensitive to concentration parameter.  The high
concentration tail of the log-normal concentration distribution
contributes to its higher mean flux, while the more extended low
concentration range from the Gaussian distribution manifests itself
into a broader distribution of fluxes at each particular angular
Galacto-centric distance.

It should also be emphasized that the shown distribution functions are
uncorrelated, while in reality, because a single concentration must be
defined for the Halo, there are correlations between the flux at
adjacent angular bins, and anti-correlations between angular values
close to zero and 180 degrees.  A higher concentration halo has
relatively more mass near the center and less near the outer parts.
Thus, highly concentrated halos will have higher fluxes with respect
to the distribution function toward the Galactic center, and will have
relatively smaller fluxes toward the Galactic anti-center.

 We now quantify the expected angular dependence of the peak and width
 of the flux probability distribution function.  The peak of the flux
 distribution is expected to be smaller at high angular distances from
 the Galactic center. This is a natural consequence of the centrally
 concentrated spherical distribution of dark matter in a halo, and
 enables a measure of the underlying density profile of the halo. In
 Fig.~\ref{fig:MWFluxprop1}, we show the angular dependence of the
 peak flux. A good fit (within few percent) to this function is
 obtained by a double power-law as:
\begin{equation} 
\ln \bar{\Phi} = a_1 + a_2  \ln \psi + a_3 \ln ( a_4 +
\psi) 
\label{eqnpeak}
\end{equation}
where the parameters $a_1$, $a_2$, $a_3$ and $a_4$ are given in
Table~\ref{table:table}.  The width of the distribution is smaller at
large angles from the Galactic center. This is to be expected because
the effects of concentration are more apparent in the inner regions of
the halo. At radial distances $r \gg r_{-2}$, the changes in the dark
matter density due to different halo concentration values are smaller
and therefore the flux distribution is narrower. In
Fig.~\ref{fig:MWFluxprop2}, we show the expected angular dependence of
the width of the flux distribution. We find that a function of the
form
\begin{equation} 
\sigma = a_5  \ln \psi + a_6
\label{eqnangle}
\end{equation} 
is a good (within few percent) fit to the angular dependence of the
width of the distribution function. The quantities $ a_5$ and $a_6$
are given in Table~\ref{table:table}.  It is important to note here
that for large radii ($r \gg r_{-2}$) the halo density PDF that we
measured in \S~\ref{subsec:density} is larger than inferred from the
concentration scatter, which implies that the values of flux
uncertainty at large Galacto-centric angles may be larger than our
estimates.  However, because the Solar Radius is well within $r_{-2}$,
the flux should be dominated by the density distribution within
$r_{-2}$, even toward the Galactic anti-center, so effects of large
radii scatter on the flux PDF should be small.

\begin{table}
\begin{center}
\begin{tabular}{c|c|c|c|c|c|c}
& $a_1 $& $a_2 $ & $a_3$ & $a_4$ & $a_5$ & $a_6 $ \\
\hline
${\rm Gaussian}$ & -13.70 & -0.39 & -1.95 & 3.27 & -0.12 & 0.98 \\
\hline
${\rm Log-normal}$ & -13.48 & -0.40 & -1.96 & 3.22 & -0.11 & 0.87 \\
\hline
\end{tabular} 
\caption{Fitting parameters of Eq. \ref{eqnpeak} \& \ref{eqnangle} for
  the peak and width of the gamma-ray annihilation flux distribution
  function along a Galactic line of sight.}
\label{table:table}
\end{center}  
\end{table}

\subsection{Other applications}

\subsubsection{Cosmological Uncertainty of the Local Dark Matter Density}
Applying our results, as before, to a Milky Way mass of
$M_{200}=1.8\times10^{12}\msun$, with an ($\alpha = 0.1645$) Einasto
profile of concentration of $c_{200}=5.97$ (using the
concentration-mass relation of \citealt{maccioconcs}), and a Solar
radius of 8.5 kpc implies a Solar radius total matter density of
$0.210^{+0.42}_{-0.45} {\rm GeV} \, {\rm cm}^{-3}$ if the
concentration PDF is assumed to be Gaussian with width proportional to
our values of $\sigma_c/c=0.283$.  Assumptions of a log-normal
distribution of width $\sigma_{LN} = 0.121$ results in only minor
changes for a local density range of $0.210^{+0.047}_{-0.039} {\rm
  GeV} \, {\rm cm}^{-3}$, which can be compared with several
observational estimates.  \cite{bergstrom} find an allowed range of
[0.2-0.8] ${ \rm GeV} \, {\rm cm}^{-3}$, and a more recent work by
\cite{weber} find an acceptable range of [0.2-0.4] ${\rm GeV} \, {\rm
  cm}^{-3}$.  However, tighter constraints are found by \cite{widrow}
and \cite{catena}, who utilize a variety of dynamical observables to
estimate, respectively, $0.304\pm0.053$ ${\rm GeV}\, {\rm cm}^{-3}$,
and $0.385\pm 0.027$ ${\rm GeV} \, {\rm cm}^{-3}$ for the dark matter
density.  These estimates are somewhat larger than our cosmological
range, which hints at the possibility that the Solar radius dark
matter density has been enhanced by ``adiabatic contraction'' in
response to baryon cooling.

\subsubsection{Implications for Halo Stacking}

Our results have implications for many astrophysical applications that
depend upon the mass distribution within halos.  One such example is
the technique of ``stacking''halos to improve signal to noise,
commonly employed in simulations and observations.  In one such
application, large numbers of simulated halo density profiles are
stacked to measure the mean density profile to high precision (\eg
\citealt{gaoconcs}; \citealt{hayashiwhite}).  From an observational
perspective, stacking many halos of similar mass greatly reduces the
noise in, for example, weak lensing determinations of halo mass
profiles and concentrations (\eg \citealt{mandelbaum08};
\citealt{mandelbaum09}; Sheldon \etal 2009ab).
 
\begin{figure}
  \includegraphics[width=.5\textwidth]{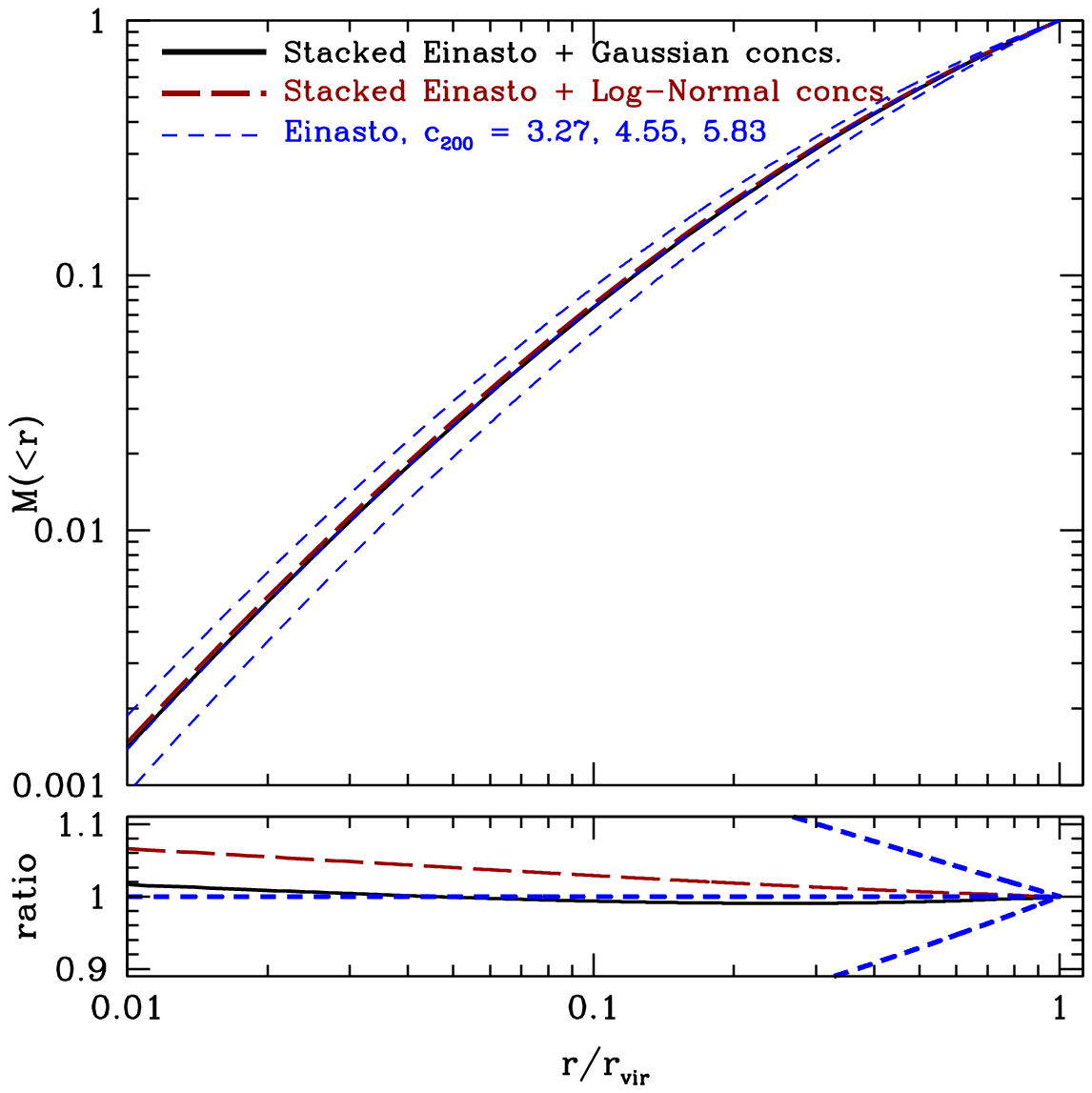}
  \caption{ Enclosed mass in spherical shells for mock stacked halo
    samples with Einasto profile at $c_{200}=4.55$ plus Gaussian
    concentration distribution and Log-normal concentration
    distribution.  $c_{200}=3.27$ and $c_{200}=5.83$ denote the
    $1-\sigma$ range in concentration distribution.  Bottom panel
    shows the ratio of each mass profile with respect to mean
    concentration, $c_{200}=4.55$, Einasto profile.  Level of
    agreement with mean concentration indicates robustness of $M<r$ in
    stacked halo samples against bias caused by halo-to-halo scatter
    in concentrations.  }
  \label{mltr}
\end{figure}

Our results support the viability of halo ``stacking''.  Due to the
fact that the distribution of densities at fixed radius is Gaussian
and the distribution of concentrations is also Gaussian, a stacked
density profile is indeed an accurate representation of the median
profile.  This is particularly convenient in that it allows the mean
halo profile to be parameterized into an analytic form without bias,
and allows unbiased stacking of observational mass profiles.  In
Fig.~\ref{mltr}, we have verified that the cumulative cluster mass
distribution also remains unbiased.  We compare the Einasto form of a
3-dimensional enclosed mass profile of $c_{200}=4.55$ and $\alpha =
0.19$ with the same quantity for a mock stacked halo drawn from a
Gaussian distribution of concentrations of mean $c_{200}=4.55$ and
scatter $\sigma_G = 0.283 c_{200}$, and find agreement to better than
$1\%$, except within a few percent r$_{vir}$ where differences
approach $2\%$.  This implies that mass profiles determined by lensing
studies should be free from biases associated with halo stacking.
Cosmological simulations have been used to demonstrate that accurate
three dimensional mass profiles can be constructed from stacked shear
signals (\eg \citealt{johnston}); our work shows that any potential
systematic bias related to the distribution of halo concentrations or
densities will be negligible for a gaussian concentration PDF, even
for future precision surveys.  If, instead, the density distribution
had been log-normal, then a stacked halo would have been biased toward
higher cumulative masses at small radii, by more than $6\%$ at
$1\%$r$_{vir}$ in our test case.

\section{Limitations of this work}
\label{sec:limitations}

In this work, we do not consider other potential important factors to
the annihilation rate, which could possibly be dominant over the
halo-to-halo scatter associated with hierarchical structure formation
(via the distribution of concentrations) that we have examined.  Among
them is the amount of substructure in small subhalos and streams (the
``clumpy'' dark matter component), whose contribution can boost the
annihilation rate relative to that of a smooth halo, and will add to
the uncertainty in the local dark matter density (\eg
\citealt{kamionkowski,2010PhRvD..81d3532K}, \citealt{vogelsberger}).
We also ignore any gravitational coupling that the differential
evolution of baryonic halo component may have on the dark matter halo
structure.  Baryon influences may include gas cooling; this could
cause the dark matter halo to respond to the deeper potential by
``adiabatic contraction''(\citealt{blumenthal}).  However, strong
stellar or AGN feedback could instead lead to shallowing of the dark
matter potential (\eg \citealt{duffybaryons}).  Although these effects
may be important, they are beyond the scope of this study.

Our measurements of the cosmological distribution of halo
concentrations, densities, and other quantities utilized only clusters
from the simulation (because those are the best resolved halos).  Our
application toward annihilation rates in the Galaxy relies upon the
assumption that the behavior of the halo-to-halo profile scatter is
similar for galaxies and clusters, namely that the probability
distribution of the mean density in radial shells is always described
by the Einasto profile with a Gaussian distribution in halo
concentrations.  The assertion that the distribution of halo
concentration remains universally Gaussian, while speculative, is
supported by the fact that the logarithmic width and shape of the
concentration distribution has weak or no mass or redshift dependence
(\eg \citealt{bullock}; \citealt{netoconcs}; \citealt{gaoconcs};
\citealt{maccio}).  This is expected from the self-similar scatter in
formation time with mass and the close correlation between formation
time and concentration (\citealt{wechsler}).  Additionally,
\cite{knollmann} used scale-free simulations to show that the scatter
in halo profile concentration and density slope has little dependence
on matter power spectral index (which varies with halo mass) over a
range bracketing well beyond the effective spectral indices of
clusters and galaxies.  Future studies are warranted utilizing a wider
range in halo masses to determine whether the distribution of
concentrations is universally Gaussian.

\section{Conclusions}
\label{sec:conclusions}

The probability distribution function of dark matter within halos, as
we have explored in this work, provides some basis for interpreting
both indirect and direct dark matter detection experiments in a
cosmological context.  Constraints upon the dark matter density,
particle mass, or the self-annihilation cross section depend on the
probability distribution function of dark matter.

Our results indicate that halo concentration is the primary
cosmological contributor to the dark matter PDF.  This implies a
particular correlation between the local dark matter density, relevant
for direct detection efforts, and the dark matter density in the
direction of the Galactic center (and elsewhere), applicable to
indirect detection experiments.  The effect of halo concentration
should thus be a crucial factor in verifying the consistency of dark
matter density constraints made from multiple dark matter detection
techniques.  Ultimately, dark matter signals might be able to test the
validity of the $\Lambda$CDM cosmological model through estimates of
the dark density at differing locations within the Milky Way halo, and
perhaps also within other halos.

\section{acknowledgments}
This work was partially supported by the DOE through the IGPP, the
LDRD-DR and the LDRD-ER programs at LANL.  We thank Carlos Frenk, Tom
Theuns, Salman Habib, Katrin Heitmann, and Zarija Luki{\'c for helpful
  discussions.  DR thanks KITP for its hospitality where portions of
  this work were completed.  LG acknowledges support from the
  one-hundred-talents program of the Chinese academy of science (CAS),
  the National basic research program of China (program 973 under
  grant No. 2009CB24901), {\small NSFC} grants (Nos. 10973018) and an
  STFC Advanced Fellowship, as well as the hospitality of the
  Institute for Computational Cosmology in Durham, UK. We thank the
  Virgo Consortium for kindly allowing us use of the Millennium
  simulation.  We are grateful to the anonymous referee for insightful
  suggestions.

{}

\label{lastpage}
\end{document}